\pdfoutput=1
\documentclass[12pt]{iopart}
\usepackage[sort&compress,square,numbers]{natbib} 

\usepackage{multicol}
\usepackage{multirow}

\usepackage{graphicx}
\usepackage[usestackEOL]{stackengine}

\usepackage{subcaption}

\usepackage{color}

\begin{document}

\title[Towards coarse-grained elasticity of 2D COFs]{Towards coarse-grained elasticity of single-layer Covalent Organic Frameworks}

\author{Alexander Croy$^{1}$, Antonios Raptakis$^{2,3}$, David Bodesheim$^{2}$, Arezoo Dianat$^{2}$, and Gianaurelio Cuniberti$^{2,4}$}
\address{$^{1}\space $ Institute of Physical Chemistry, Friedrich Schiller University Jena, 07737 Jena, Germany}
\address{$^{2}\space $ Institute for Materials Science and Max Bergmann Center of Biomaterials, TU Dresden, 01062 Dresden, Germany}
\address{$^{3}\space $ Max Planck Institute for the Physics of Complex Systems, 01187 Dresden, Germany}
\address{$^{4}\space $ Dresden Center for Computational Materials Science (DCMS), TU Dresden, 01062 Dresden, Germany}
\ead{alexander.croy@uni-jena.de}

\begin{abstract}
Two-dimensional covalent organic frameworks (2D COFs) are an interesting class of 2D materials since their reticular synthesis allows the tailored design of structures and functionalities. For many of their applications the mechanical stability and performance is an important aspect. Here, we use a computational approach involving a density-functional based tight-binding method to calculate the in-plane elastic properties of about 40 COFs with a honeycomb lattice. Based on those calculations, we develop two coarse-grained descriptions: one based on a spring network and the second using a network of elastic beams. The models allow us to connect the COF force constants to the molecular force constants of the linker molecules and thus enable an efficient description of elastic deformations. To illustrate this aspect, we calculate the deformation energy of different COFs containing the equivalent of a Stone-Wales defect and find very good agreement with the coarse-grained description.
\end{abstract}
\maketitle

\section{Introduction}\label{sec:intro}
Two-dimensional covalent organic frameworks (2D COFs) are nanostructured porous crystals and have shown considerable potential for applications in many fields \cite{Li2020, Rodriguez2020,Colson2013,Diercks2017,Kim2020,Jiang2020}. They consist of networks formed by covalently linked organic molecules made from light elements
such as carbon, nitrogen, boron and oxygen. This specific molecular architecture gives them a high mechanical, thermal and chemical stability \cite{Ding2013}. The underlying reticular chemistry allows it to combine different building blocks to obtain specific topologies and properties \cite{Colson2013,Diercks2017,Kim2020,Jiang2020}. Compared to their inorganic cousins, e.g.\ TMDCs, this leads to a much larger variety of possible materials but also requires more sophisticated synthesis strategies \cite{Zwaneveld2008,Dienstmaier2011,Ortega2021,Wang2019_Dresden, Sahabudeen2016,Dai2016,Dong2018,Shao2018,Zhang2019,Liu_Kejun2019,Zeng2022}.

Although many properties of 2D COFs, and in particular structural and elastic properties \cite{Zhou2010,Duong2019,Ziogos2020}, can be calculated using a range of different methods, the large number of atoms in the unit-cell and the
combinatorial wealth of possible COFs still poses a challenge. Moreover, for COFs containing defects, like single 5+7 defects \cite{Xu2014} or extended grain boundaries \cite{Qi2020,Castano21}, a full atomistic description quickly becomes unfeasible and coarse-grained treatments are unavoidable. Recently, we have proposed one such approach in the context of elasticity where the COFs are treated as an effective spring network and the spring constants are obtained directly from the individual building blocks \cite{Raptakis2021}. For several COFs with square-lattice geometry, we could predict the 2D bulk-modulus from the molecular spring constants.
\begin{figure*}[t!]
    \centering
    \includegraphics[width=\textwidth]{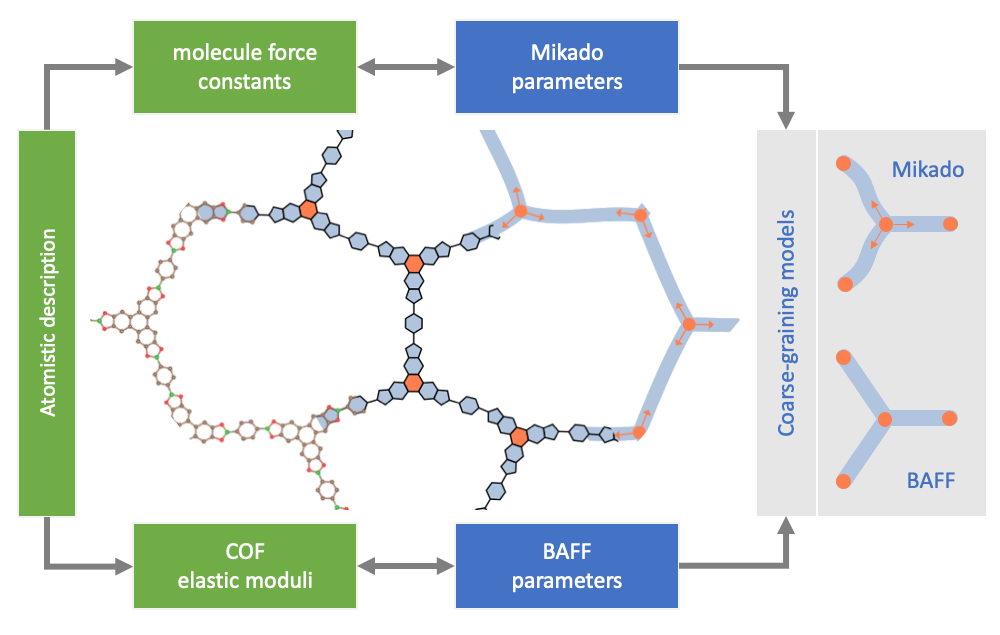}
    \caption {Schematic view of the coarse-graining approach: From the atomistic description of the COF the elastic moduli are computed and used to obtain the parameters of the bond and angle force field (BAFF) which uses only the positions of the cores (orange areas) to get lengths of the linkers (blue areas) and angles between them. Additionally, the force constants of linker molecules are used to determine the parameters of the Mikado model. Here, the linker molecules are represented by elastic beams which are cross-linked at the cores. The latter are considered as rigid molecules having a position and an orientation (arrows).
    In the center a Tp DB-COF (or COF-5) with a Stone-Wales defect is shown as an example. Carbon atoms are colored brown, oxygen red and boron green. Hydrogen atoms are not shown for convenience.}
	\label{fig:coarse_graining}
\end{figure*}

In this article we focus on COFs with honeycomb lattices since their material properties are expected to be isotropic \cite{Landau1986}. In addition to the bulk modulus we also consider the shear modulus and thus achieve a complete characterization of the in-plane elasticity of those materials. Using a description in terms of a coarse-grained spring network which includes bond-stretching and angle-bending contributions, we can connect the effective force constants of the network to the elastic moduli (see Fig.\ \ref{fig:coarse_graining}). 
In order to better account for the flexibility of the linker molecules,
we develop a second model which is based on treating them as 2D elastic beams. This model, which is inspired by the well-known Mikado model \cite{Wilhelm2003,Head2003} and beam-network models \cite{Lagnese1993,Phani2006}, allows it to understand the relation of bulk and shear moduli with lattice constant of the COFs and provides a connection of the COF elasticity with the molecular properties. Both coarse-grained descriptions can also be applied to structures with defects. To validate and benchmark the models we performed calculations with the density functional-based tight-binding method (DFTB) \cite{Frauenheim.1998,Elstner.2007} for about forty COFs and their building-blocks. In three cases we introduce the equivalent of a Stone-Wales defect \cite{Stone1986} into the crystal and compute the resulting deformation energies.

\begin{figure}[t!]
    \centering
    \begin{subfigure}[t][][c]{0.53\textwidth}
        \caption{}
        \includegraphics[width=\textwidth]{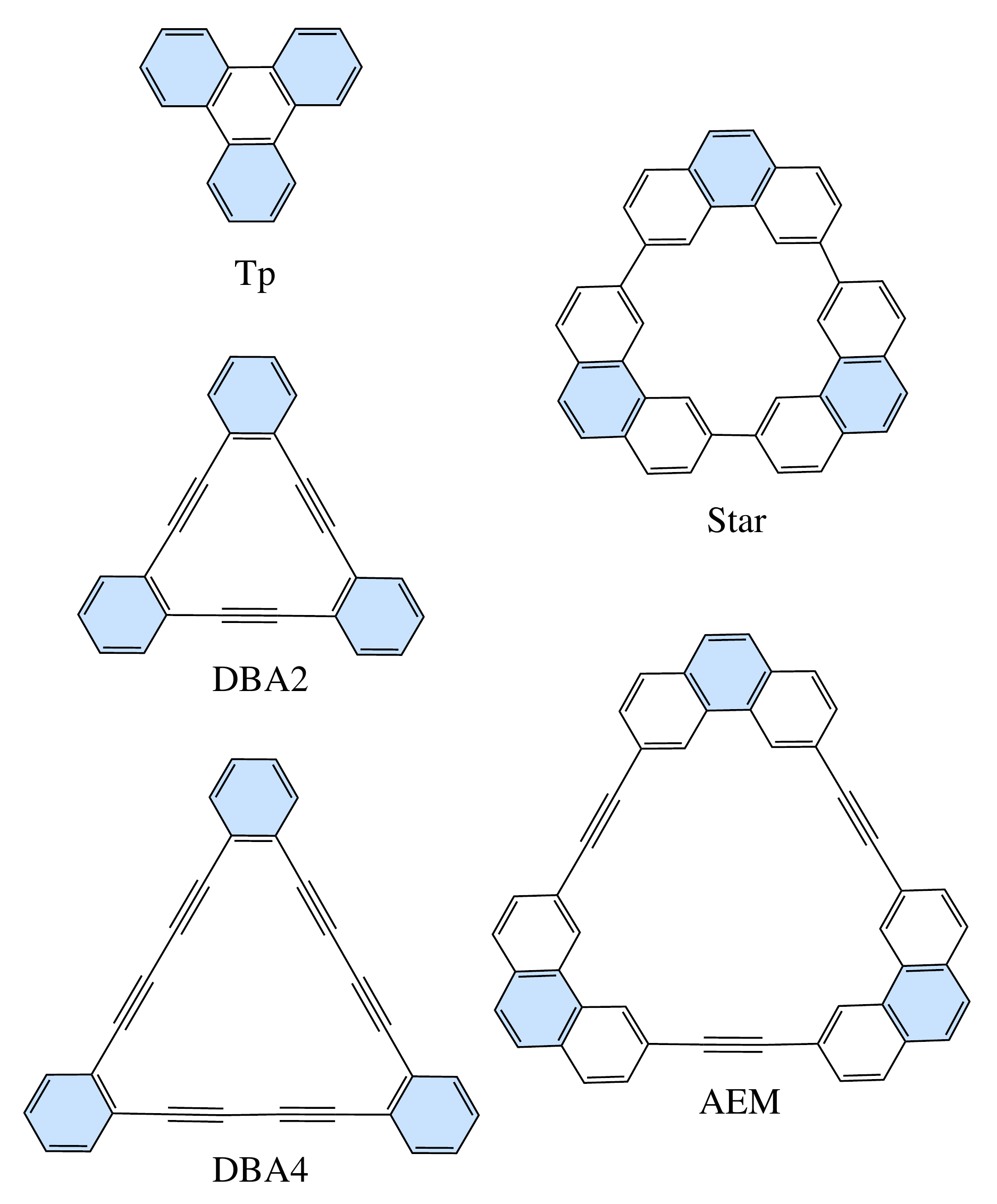}
    \end{subfigure}
    \begin{subfigure}[t]{0.45\textwidth}
        \caption{}
        \includegraphics[width=\textwidth]{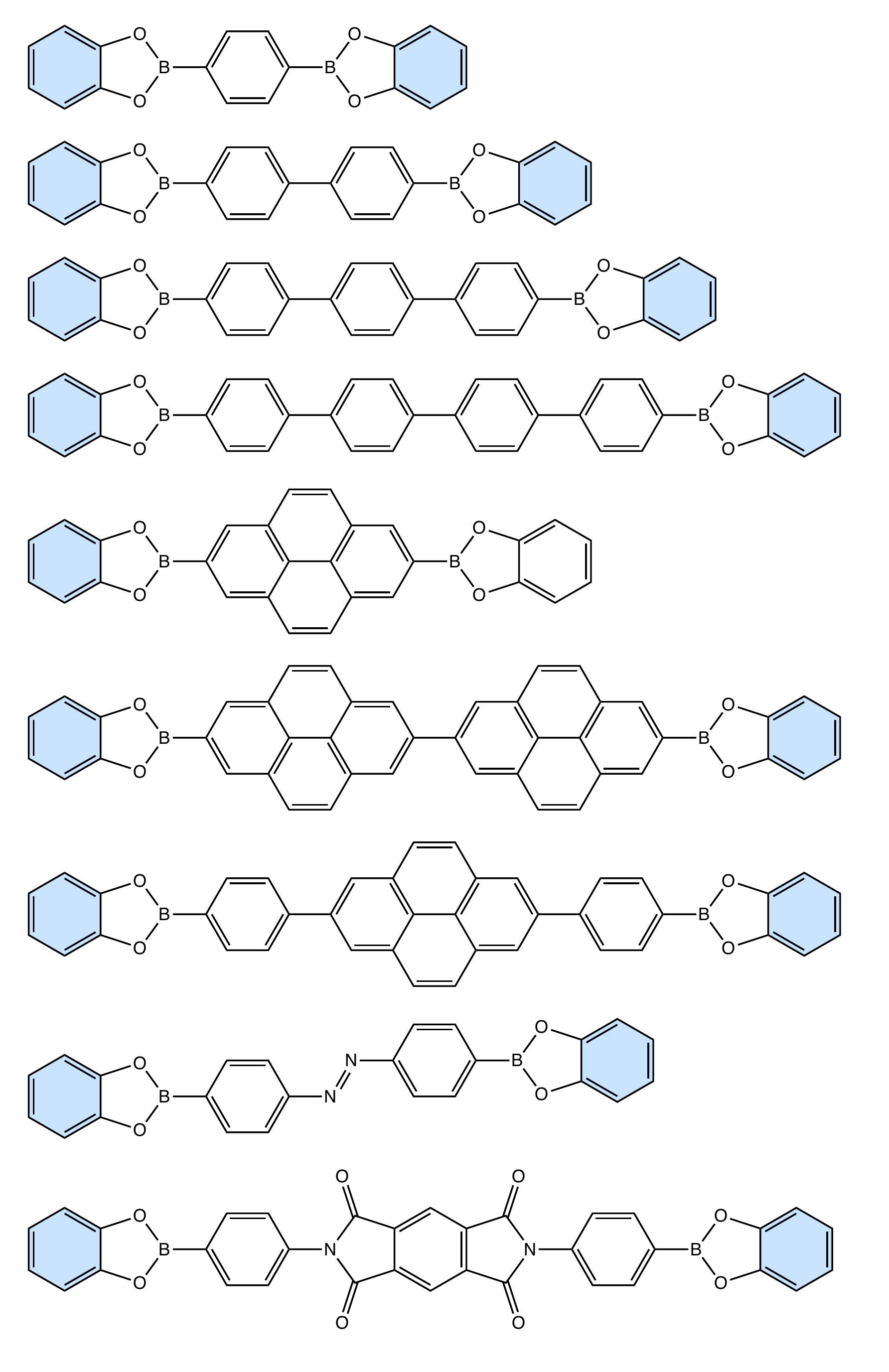}
    \end{subfigure}\\
    \begin{subfigure}[b]{0.9\textwidth}
        \caption{}
        \includegraphics[width=\textwidth]{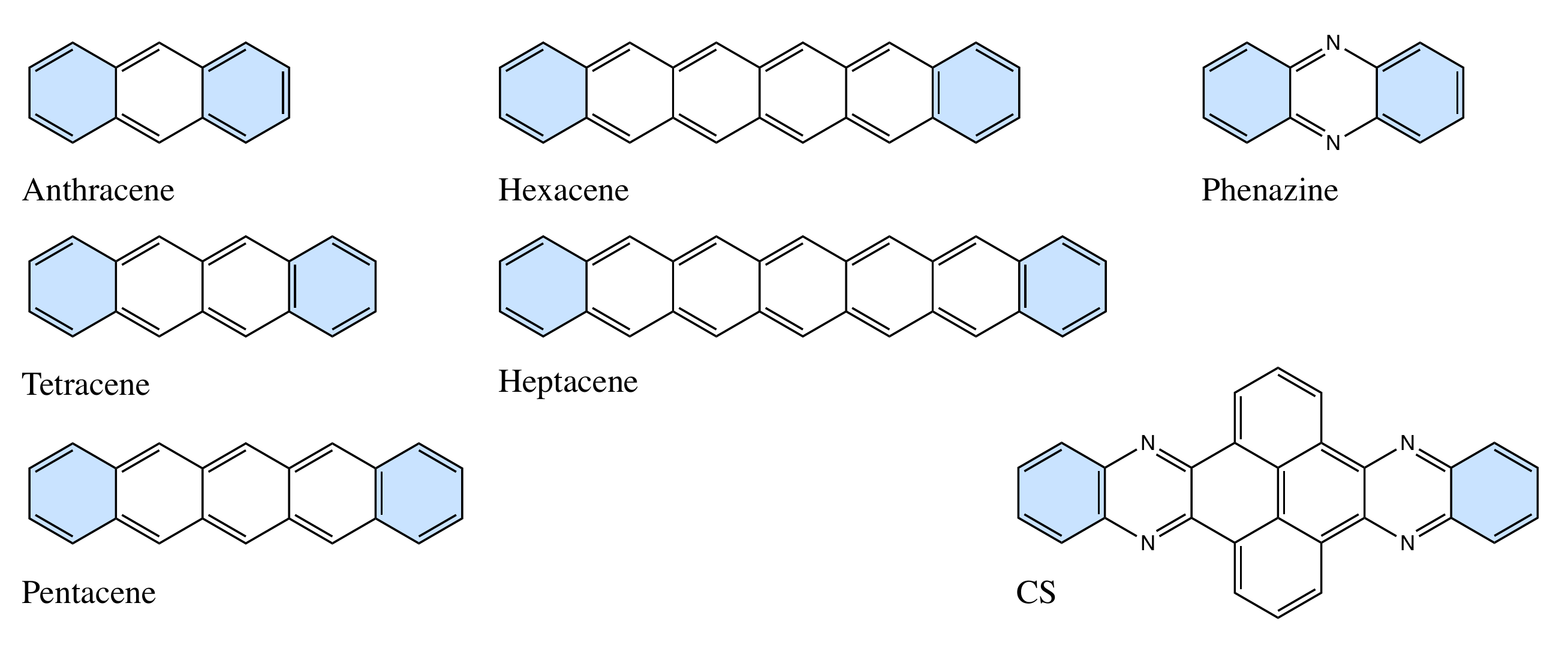}     
    \end{subfigure}
    \caption {Cores and linker molecules. The blue colored phenyl rings are shared between core and linkers. (a) 
    Five cores with $C_3$ symmetry are used in this work: Triphenylene (Tp), Star, AEM, DBA2, and DBA4. (b) Linker molecules containing boronate ester, from top to bottom: DB 1phenyl, DB 2phenyls, DB 3phenyls, DB 4phenyls, DB 1pyrene, DB 2pyrenes, DB phenyl pyrene phenyl, DB Azobenzene, and DB ANDI. (c) Linker molecules only used with Tp core: Anthracene, Tetracene, Pentacene, Hexacene, Heptacene, Phenazine, and pyrene-fused tetraazahexacene (CS).
    }
	\label{fig:models}
\end{figure}
The specific COFs studied in this work include five molecular cores with $C_3$ symmetry and different linear linker molecules. In Fig.\ \ref{fig:models} the molecular building-blocks are shown together with the respective linker molecule used for the computational model. It should be noted that those linkers are not necessarily identical to the building-blocks used for the polymerization. Instead, our nomenclature reflects the structure of the resulting COFs. We consistently denote them as \textit{core}--\textit{linker}--COF, for example, COF-10 is denoted as Tp--DB 2phenyls--COF. The COFs under consideration include Tp--DB 1phenyl--COF (known as COF-5) \cite{COF5}, Star-COF\ \cite{Feng2013}, DBA2- and DBA4-COF\ \cite{Baldwin2015}, and AEM-COF\ \cite{Yang2015_AEM}. In order to obtain linkers with a larger variability of lengths, we keep the molecular structure but change the number of phenyls or pyrenes in the linkers \cite{Cote2007,Rager2017}, as shown in Fig.\ \ref{fig:models}(b). Additionally, we consider linkers based on fused acenes, phenazine and pyrene-fused tetraazahexacene \cite{Kou2011,Pham2019,Guo2013}.

The article is structured as follows. In section \ref{sec:theory}, the two coarse-grained models are introduced and expressions for the elastic moduli and force constants are derived. In the next section, the computational details are provided. Section \ref{sec:results} contains the presentation and discussion of the results. Conclusions are provided in the last section.

\section{Elasticity of honeycomb 2D COFs}\label{sec:theory}
\subsection{Elastic moduli of coarse-grained spring networks}\label{sec:spring_network}
A straight-forward coarse-graining description of a COF consists in treating the cores as sites of a lattice. Those sites have only translational degrees of freedom and are mutually connected via the linkers, which are not further characterized. More specifically, we consider a honeycomb lattice with interactions between neighboring sites $i$ and $j$ (stretching contribution) and between neighboring bonds (angle bending contribution). The elastic energy of the resulting Bond and Angle Force Field (BAFF) reads \cite{Perebeinos2009}
\begin{eqnarray}
    \mathcal{H}_{\rm total} &={}& \sum_{i,j} \frac{\beta_r}{\ell_0^2} (\ell_{ij} - \ell_0)^2
     + \sum_{i,j<k} \beta_c (\cos(\phi_{jik}) - \cos(\phi_{0}))^2 \nonumber\\
&={}&\frac{1}{2}\sum_{b} \frac{4\beta_r}{\ell_0^2} (\ell_{b} - \ell_0)^2
     + \sum_{b,b'} \frac{3\beta_c}{8} (\phi_{bb'} - \phi_{0})^2\;, \label{eq:cg_energy_honeycomb}
\end{eqnarray}
where $\ell_{ij}=|\vec{R}_i-\vec{R}_j|$ is the distance between sites $i$ and $j$,
$\phi_{jik}$ is the angle between the bond-vectors $\vec{R}_{ij} = \vec{R}_j-\vec{R}_i$ and $\vec{R}_{ik} = \vec{R}_k-\vec{R}_i$, and $\ell_0$ and $\phi_{0}$ are the equilibrium values. In the second line, summations are over bonds, $b$ and $b'$, instead of sites and the bending term was expanded for angles close to the equilibrium.

In terms of the force constants $\beta_r$ and $\beta_c$ the 2D bulk and shear modulus are \cite{Perebeinos2009}
\numparts\label{eq:bulk_shear}
\begin{eqnarray}
    B_{2D} ={}& 
    \frac{ 4\beta_r}{2\sqrt{3}\ell_0^2} 
    = \frac{k}{2\sqrt{3}}\;,\\
    \mu_{2D} ={}& 
    {3}\sqrt{3} \frac{ k \beta_c/\ell_0^2}{k + 9\beta_c/\ell_0^2}
=
    \frac{1}{\sqrt{3}} \frac{ k \kappa}{k + \kappa} = \frac{1}{\sqrt{3}} \left( \frac{1}{k} + \frac{1}{\kappa}\right)^{-1}\;.
\end{eqnarray}
\endnumparts
The bulk modulus $B_{2D}$ only depends on the stretching force-constant $k=4\beta_{r}/\ell_0^2$, while the shear modulus $\mu_{2D}$ is determined by $k$ and the effective angle-bending constant $\kappa=9\beta_c/\ell_0^2$.

Due to the honeycomb symmetry, all in-plane elastic constants can be expressed in terms of $B_{2D}$ and $\mu_{2D}$ \cite{Landau1986}. In particular, the non-vanishing elements of the stiffness tensor are (see also supporting information)
\numparts
\begin{eqnarray}
    C_{11} ={}& C_{22} = B_{2D} + \mu_{2D}\;,\\
    C_{12} ={}& C_{21} = B_{2D} - \mu_{2D}\;,\\
    C_{66} ={}& (C_{11} - C_{12})/2 = \mu_{2D}\;.
\end{eqnarray}
\endnumparts
Finally, the Young modulus $Y_{2D}$ and the Poisson ratio $\nu_{2D}$ can be obtained via
\begin{equation}
    Y_{2D} = \frac{4 B_{2D} \mu_{2D} }{B_{2D} + \mu_{2D}}\quad{\rm and}\quad
    \nu_{2D} = \frac{B_{2D} - \mu_{2D}}{B_{2D} + \mu_{2D}}\;.
\end{equation}
Note, that the energy given by Eq.\ \eref{eq:cg_energy_honeycomb} does not contain terms which penalize out of plane bending and therefore the bending rigidities vanish within this model.

\subsection{Stretching and bending moduli of semi-flexible polymers}\label{sec:semi_flex}
Two-dimensional COFs are basically cross-linked semiflexible polymer networks where the cross-links are located on a 2D surface. A very successful model of random networks is the   
Mikado model \cite{Wilhelm2003,Head2003,Head2003b}, where it is assumed that the network is formed by fibers which are cross-linked at the intersection of two such fibers (cf.\ Fig.\ \ref{fig:coarse_graining}). The individual fibers are modeled as elastic beams and their energy under deformation is given by:
\begin{equation}
    \mathcal{H}_{\rm fiber} = \int_0^{\ell_0} ds \left\{ \frac{c_{\rm stretch}}{2} \left(\frac{\partial u_{\|}(s)}{\partial s}\right)^2 + 
    \frac{c_{\rm bend}}{2}\left(\frac{\partial^2 u_{\perp}(s)}{\partial s^2}\right)^2\right\}\;,
\end{equation}
where $s$ denotes the position along the fiber axis and $u_{\|}$ and $u_{\perp}$ are the tangential and transverse displacements with respect to the fiber axis. The first contribution describes the stretching energy, which results from changing the length of the fiber, while the second term is the bending energy. The two corresponding elastic moduli are $c_{\rm stretch}$ and $c_{\rm bend}$. For an elastic beam the two moduli are related, but here we treat them as independent molecule-specific constants. More details can be found in the supporting information.

For a homogeneous elongation or compression of the fiber, i.e.\ $u_{\|}(s) = \frac{\Delta\ell}{\ell_0} s$, the stretching energy becomes:
\begin{equation}\label{eq:fiber_stretch_energy}
    \mathcal{H}_{\rm stretch} =  \frac{1}{2} c_{\rm stretch} \ell_0 \left(\frac{\Delta \ell}{\ell_0}\right)^2
     =  \frac{1}{2} \frac{ c_{\rm stretch} }{{\ell_0}}  {\Delta \ell^2}\;,
\end{equation}
which corresponds to a Hookean spring with spring constant $k_{\rm fiber}=c_{\rm stretch}/\ell_0$. The nature of the bending deformation will depend on the cross-linking, which determines the boundary conditions for $u_{\perp}$ at the end-points: $u_{\perp}(0)$, $u_{\perp}(\ell_0)$, $\varphi(0)$, and $\varphi(\ell_0)$. The latter two denote the derivative of $u_{\perp}$, i.e., $\varphi(s) = \partial u_{\perp}/\partial s$. In terms of those boundary values, the transverse displacement is
\begin{eqnarray}
    u_{\perp}(s) ={}& \left( \frac{\varphi(0) + \varphi(\ell_0)}{\ell_0^2} - 2\frac{ u_\perp(\ell_0)-u_\perp(0)}{\ell_0^3}\right) s^3 \nonumber\\ 
    {}&+ \left( -\frac{2\varphi(0) + \varphi(\ell_0)}{\ell_0} + 3\frac{ u_\perp(\ell_0)-u_\perp(0)}{\ell_0^2}\right) s^2 \nonumber\\
    {}& + \varphi(0) s + u_\perp(0)\;.
\end{eqnarray}
This expression leads to the in-plane bending energy of the fiber as follows:
\begin{eqnarray}
    \mathcal{H}_{\rm bend} ={}&  \frac{1}{2} \frac{12 c_{\rm bend}}{\ell_0^3} \left[
    \left( u_\perp(\ell_0)-u_\perp(0) \right)^2 \right.\nonumber\\
    {}&+\frac{\ell_0^2}{3} \left(  \varphi(0)^2 + \varphi(0) \varphi(\ell_0) + \varphi(\ell_0)^2\right) \nonumber\\
    {}&\left.- \ell_0 \left( u_\perp(\ell_0)-u_\perp(0) \right)\left( \varphi(\ell_0)+\varphi(0) \right)
    \right]\;. \label{eq:fiber_bend_energy}
\end{eqnarray}
If the endpoints of the fiber are displaced relatively by $\Delta = u_\perp(\ell_0)-u_\perp(0)$ while keeping $\varphi(0)=\varphi(\ell_0)=0$, 
the effective in-plane bending force-constant becomes $\kappa_{\rm fiber} = {12 c_{\rm bend}}/{\ell_0^3}$, which scales as $\ell_0^{-3}$. If $\varphi$ at the ends is allowed to change, it is beneficial to switch to the variables 
$\psi(0) = \varphi(0) - \Delta/\ell_0$ and $\psi(\ell_0) = \varphi(\ell_0) - \Delta/\ell_0$. This leads to
\begin{equation}\label{eq:fiber_bend_psi}
    \mathcal{H}_{\rm bend} = \frac{1}{2} \frac{4 c_{\rm bend}}{\ell_0} \left[
    \psi(0)^2 + \psi(0) \psi(\ell_0) + \psi(\ell_0)^2
    \right]\;,
\end{equation}
which makes it clear that without constraints the fiber tends to a straight configuration ($\psi(0)= \psi(\ell_0)=0$) with vanishing bending energy.

\subsection{Moduli of polymer networks}\label{sec:semi_flex_network}
In polymer networks the fibers are connected at cross-links which are typically considered as rigid joints implying that they have translational \textit{and} rotational degrees of freedom, but are not deformable \cite{Lagnese1993,Phani2006}. Using the results for single fibers presented in the previous section, we can obtain the effective spring and bending constants ($k$ and $\kappa$) and thus the elastic moduli of the network. The stretching energy is simply given by the sum of individual fiber contributions (cf.\ Eq.\ \eref{eq:fiber_stretch_energy}) and $k=k_{\rm fiber}$. In order to determine the deformation energy of angle bending, we consider two straight fibers (denoted A and B) of length $\ell_0$ which are linked at a rigid joint (at $s=0$). Displacing the endpoint of one fiber (A) perpendicular to the fiber axis by $\Delta$ changes the angle between the fibers from $\phi_0$ to $\phi = \phi_0 + \Delta/\ell_0$. According to Eq.\ \eref{eq:cg_energy_honeycomb} the deformation energy is
\begin{equation}\label{eq:two_fibers_angle1}
    \mathcal{H}_{\rm angle} = \frac{3 \beta_c}{2} \left(\frac{\Delta}{\ell_0}\right)^2\;.
\end{equation}
On the other hand, adding the in-plane bending energies \eref{eq:fiber_bend_psi} for both fibers yields
\begin{eqnarray}
    \mathcal{H}_{\rm angle} ={}& \frac{2 c_{\rm bend}}{\ell_0} \left[
    \psi_A(0)^2 + \psi_A(0) \psi_A(\ell_0) + \psi_A(\ell_0)^2 \right.\nonumber\\
    {}&\left. + \psi_B(0)^2 + \psi_B(0) \psi_B(\ell_0) + \psi_B(\ell_0)^2
    \right]\;.\label{eq:two_fibers_angle2}
\end{eqnarray}
Since the fibers are linked at $s=0$, the constraint $\psi_A(0) = \psi_B(0) + \phi_0 - \phi = \psi_B(0) - \Delta/\ell_0$ holds. If the deformation is such that fiber B remains straight ($\psi_B(\ell_0)=\psi_B(0)=0$)
and $\psi_A(\ell_0)=0$, the comparison of the deformation energies 
\eref{eq:two_fibers_angle1} and \eref{eq:two_fibers_angle2} gives
\begin{equation}\label{eq:fiber_bend}
    \frac{3 \beta_c}{2} = \frac{2 c_{\rm bend}}{\ell_0}
    \quad\Longleftrightarrow\quad
    \kappa = \frac{9 \beta_c}{\ell_0^2} = \frac{12 c_{\rm bend}}{\ell_0^3} = \kappa_{\rm fiber}\;.
\end{equation}

Based on the expressions above, the total energy of the deformed polymer network is given as a sum of the individual fiber contributions:
\begin{equation}
    \mathcal{H}_{\rm total} = \sum_{b} \left\{ \frac{1}{2} k_{\rm fiber}  {\Delta \ell_b^2} 
        + \frac{\ell_0^2\kappa_{\rm fiber}}{6} \left[
    \psi_b(0)^2 + \psi_b(0) \psi_b(\ell_0) + \psi_b(\ell_0)^2
    \right] \right\}\;. \label{eq:cg_energy_Mikado}
\end{equation}
Additionally, at each joint the constraint
\begin{equation}
    \psi_b(0) - \psi_{b'}(0) = \phi_0 - \phi_{b b'}
\end{equation}
has to hold for all $n_j$ neighboring fibers $b$ and $b'$. Those $n_j-1$ independent conditions determine all $\psi$ at that joint in terms of its orientation. The corresponding orientation vector $\vec{\omega}$ is set to point along one of the fibers in the undeformed state (all $\phi_{bb'}=\phi_0$). Then, the angle between $\vec{\omega}$ and the direction of the chosen fiber $b$ determines $\psi_b$ as explained in the supplementary information.

\section{Computational details}\label{sec:computational}
In this work, we use DFTB with self-consistent charge extension (SCC) method \cite{Frauenheim.1998,Elstner.2007} to compute the elastic properties of COFs and monomers, as implemented in the DFTB+ code (version 20.2). For all calculations we use the matsci-0-3 parametrization \cite{matsci}. For the COF calculations, periodic boundary conditions were used. Single layers were optimized using the SCC method with $30~\textrm{\AA}$ distance between the layers to avoid interactions between them. The $k$-points according to Monkhorst and Pack \cite{Monkhorst} were generated by using a $k$-spacing of $0.3~\textrm{\AA}^{-1}$ where the resulting k-point grid was rounded to its next integer value. A convergence criterion for the SCC cycles of $10^{-8}~\textrm{e}$ and a maximum force component of $10^{-5}~\textrm{Hartree}/\textrm{bohr}$ was chosen. To determine the bulk and the shear moduli, the primitive cell was biaxially strained starting from the equilibrium configuration in steps of $0.2$\%, while the $z$-direction of the lattice vector was always remaining the same. The moduli were then determined from the total energy by fitting a cubic polynomial to the area-energy relations for the bulk or the strain-energy relations for the shear modulus \cite{Raptakis2021}. The deformation energy of COFs with SW defect was obtained for $3\times3$ supercells and using one k-point centered at the $\Gamma$-point during the geometry optimization with $10^{-5}~\textrm{e}$ and $10^{-5}~\textrm{Hartree}/\textrm{bohr}$ as convergence criteria.

Similarly, the molecular spring and bending constants were obtained from SCC DFTB calculations. We fixed the two outermost hydrogen and carbon atoms of the molecular linkers and relaxed the remaining atom positions. For obtaining the stretching force constant, we elongate and compress the molecule by moving the fixed atoms in steps of $0.2$\AA{} along the molecular axis. The force constant is then found from the energy-displacement curve via Eq.\ \eref{eq:fiber_stretch_energy}. The bending force constant is found by moving the fixed atoms perpendicular to the molecular axis and then using Eq.\ \eref{eq:fiber_bend_energy}.

\section{Results and Discussion}\label{sec:results}
\subsection{Bulk and shear moduli}
\begin{figure}[t!]
    \centering
    \includegraphics[width=1\textwidth]{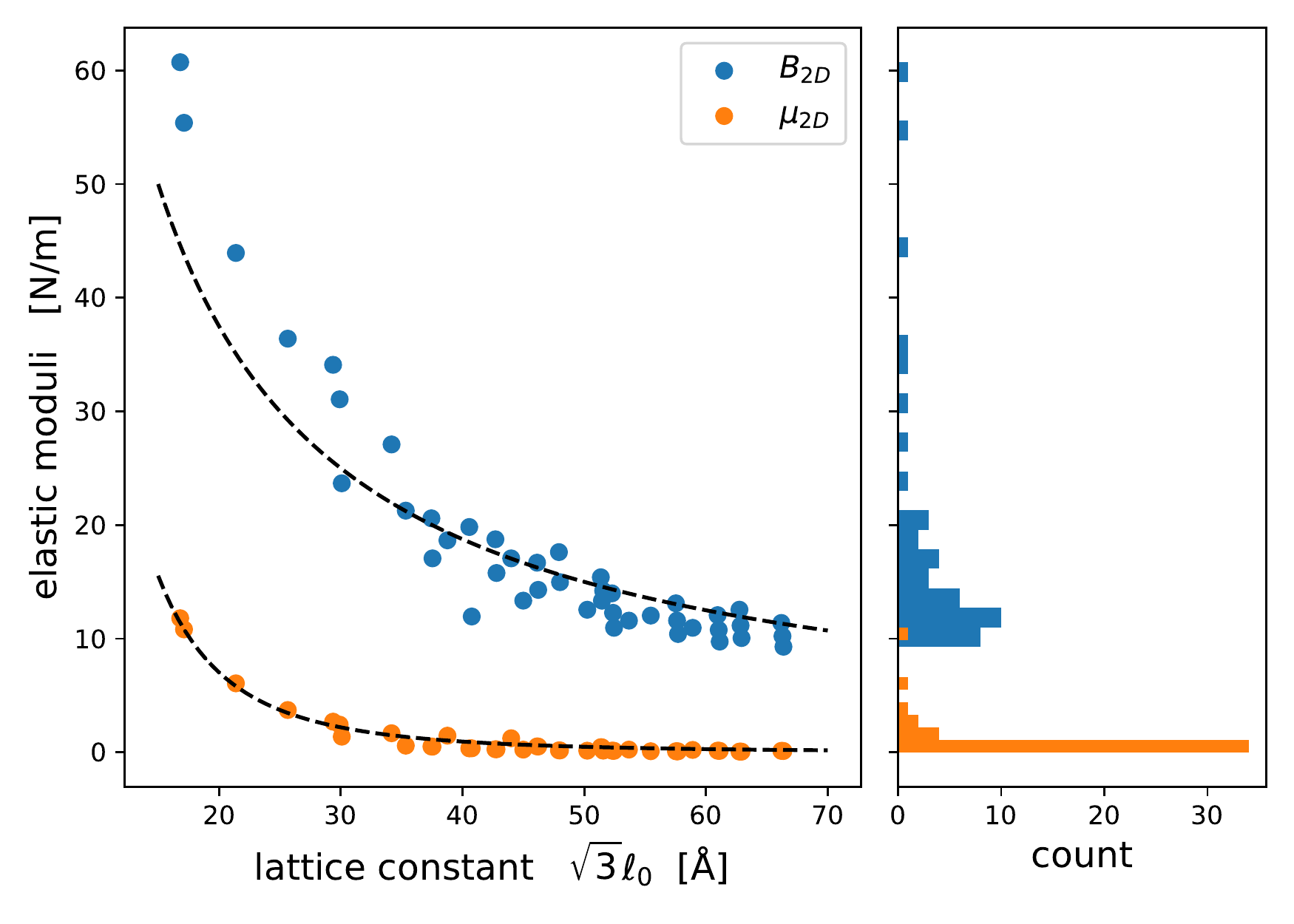}
    \caption {Lattice constant dependence of the COF elastic moduli $B_{2D}$ and $\mu_{2D}$.
    The dashed lines are guides to the eye (no fitting involved). The right panel shows the distribution of the elastic moduli for the COFs under consideration.}
	\label{fig:moduli-vs-length}
\end{figure}
The bulk and shear moduli have been calculated for the core and linker combinations shown in Fig.\ \ref{fig:models}. The respective results are summarized in Fig.\ \ref{fig:moduli-vs-length} and are listed in table \ref{tab:moduli} for a selection of COFs (see SI for the full list). Compared to graphene, the moduli of the COFs are smaller by a factor 4 to 20, but are similar to the moduli of some inorganic 2D materials. For example, the computational 2D materials database \cite{Haastrup_2018} gives\footnote{We use the reported elements of the stiffness tensor, $C_{11}, C_{22}, C_{12}$ and $C_{21}$, to calculate $B_{2D}$ and $\mu_{2D}$.} for hexagonal PbSe$_{2}$ the moduli $B_{2D}=23.89$ N/m and $\mu_{2D}=1.13$ N/m. As a general trend, the moduli decrease with increasing lattice constant. Apparently, this behavior follows a systematic trend which is indicated in Fig.\ \ref{fig:moduli-vs-length} by the dashed lines.

\strutlongstacks{T}
\begin{table*}[tb]
    \centering
    \begin{tabular}{cc|cc|cc}
    \hline
    COF Name & $a$ [Å] & $B_{2D}$ [N/m] & $\mu_{2D}$ [N/m] & $k$ [N/m] & $\kappa$ [N/m] \\
    \hline
    Tp--Anthracene   & 17.125 & 55.39 & 10.82 & 191.88 & 20.76  \\
    Tp--DB 1phenyl          & 30.083 & 23.67 & 1.38 & 82.00 & 2.46 \\
    DBA2--DB 1phenyl            & 35.339 & 21.27 & 0.59 & 73.66 & 1.03 \\
    DBA4--DB 1phenyl            & 40.556 & 19.83 & 0.35	& 68.68 &  0.61 \\
    Star--DB 1phenyl            & 38.762 & 18.65 & 1.47 & 64.62 &  2.64\\
    AEM--DB 1phenyl             & 43.998 & 17.07 & 1.24 & 59.12 & 2.22 \\
    \hline
    \end{tabular}
    \caption{DFTB results for different COFs: lattice constant ($a$), bulk and shear moduli ($B_{2D}$ and $\mu_{2D}$), and the effective spring and bending constants ($k$ and $\kappa$) obtained from Eqs.\ \eref{eq:bulk_shear}.}
    \label{tab:moduli}
\end{table*}

\begin{figure}[t!]
    \centering
    \includegraphics[width=1\textwidth]{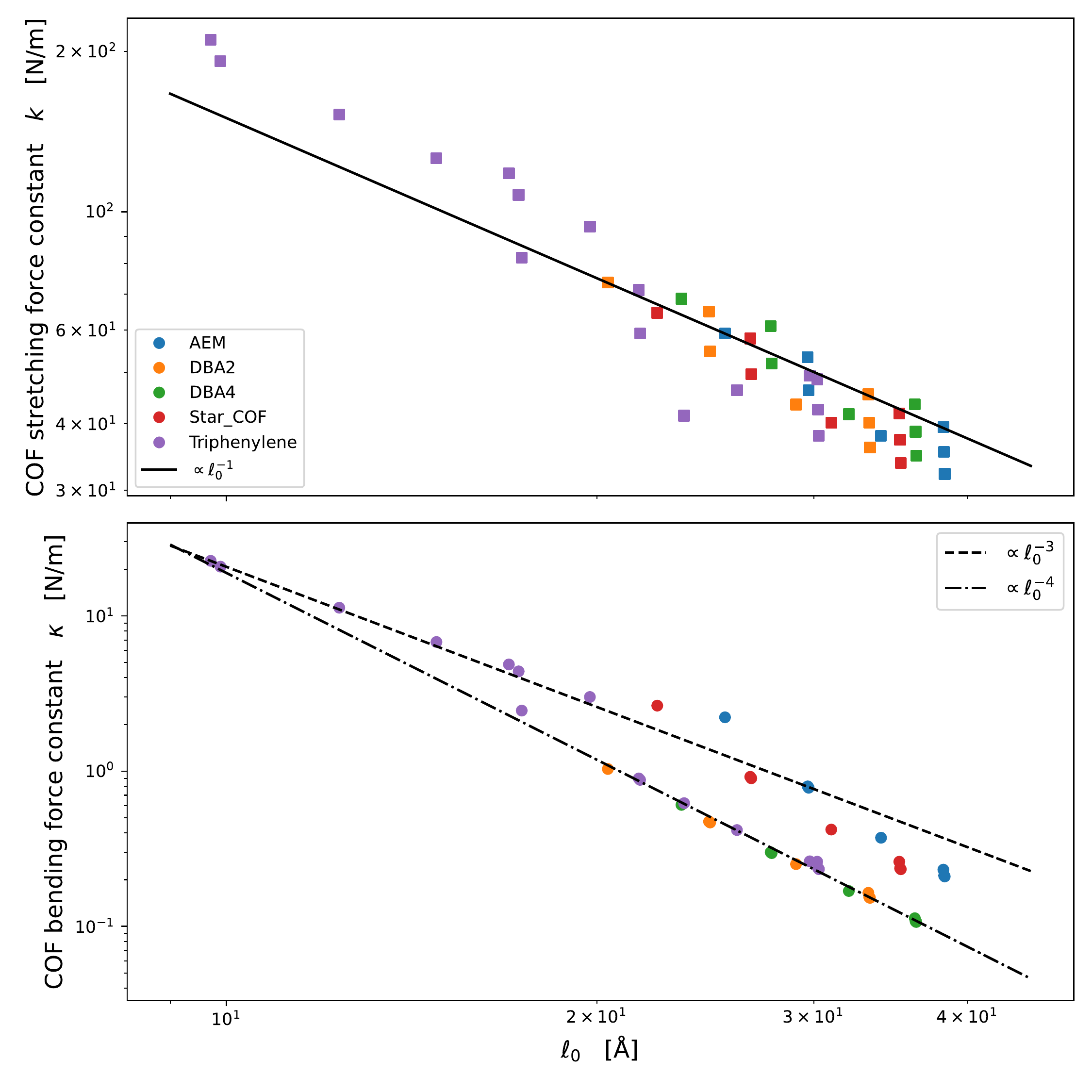}
    \caption {Length dependence of the COF force constants $k$ (top) and $\kappa$ (bottom). The length of the linker is obtained from the lattice constant $\ell_0=a/\sqrt{3}$. }
	\label{fig:fcs_vs_length}
\end{figure}
In order to understand the described behavior, we use Eqs.\ \eref{eq:bulk_shear} to obtain the effective stretching and bending force constants, $k$ and $\kappa$, for each COF. Figure \ref{fig:fcs_vs_length} shows their dependence on the length of the respective linker which is calculated from the lattice constant as $\ell_0=a/\sqrt{3}$. One can see that both force constants decrease as the length of the linker is increased. For all COFs, the stretching force constant $k$ is found to be inversely proportional to the length in accordance with the discussion of the semi-flexible fibers in Sec.\ \ref{sec:semi_flex}. On the other hand, the bending force constants are found to be either proportional to $\ell_0^{-3}$, as shown by Eq.\ \eref{eq:fiber_bend}, or are proportional to $\ell_0^{-4}$. The latter behavior is found for COFs where the cores are composed of several linked phenyl rings. The influence of this structural feature will be discussed later on.

\subsection{Elastic moduli and molecular force constants}
The results in the previous subsection show that the model of semi-flexible fibers described in Sec.\ \ref{sec:semi_flex_network} might be applicable for COFs with linear linkers. In order to investigate this question in more detail, we first look at the deformation of the linkers in the COFs for biaxial and shear deformation. We find, that in case of the biaxial deformation, the displacement is mostly in the direction of the molecular axis and there is only negligible displacement in the perpendicular direction. For the shearing, one observes the opposite behavior. A visualization of the the displacement within one linker of a Star-COF is shown in the supporting information.

Motivated by the observations of the linker displacements, we calculated the molecular stretching and bending force constants of the individual linker molecules which are given in Fig.\ \ref{fig:models}. Note that those linker molecules contain a part of the cores and are thus larger than the commonly used linker molecules. The force constants are determined from the total energy of the stretched or sheared molecule as described in the computational details.

\begin{figure}[t!]
    \centering
    \includegraphics[width=1\textwidth]{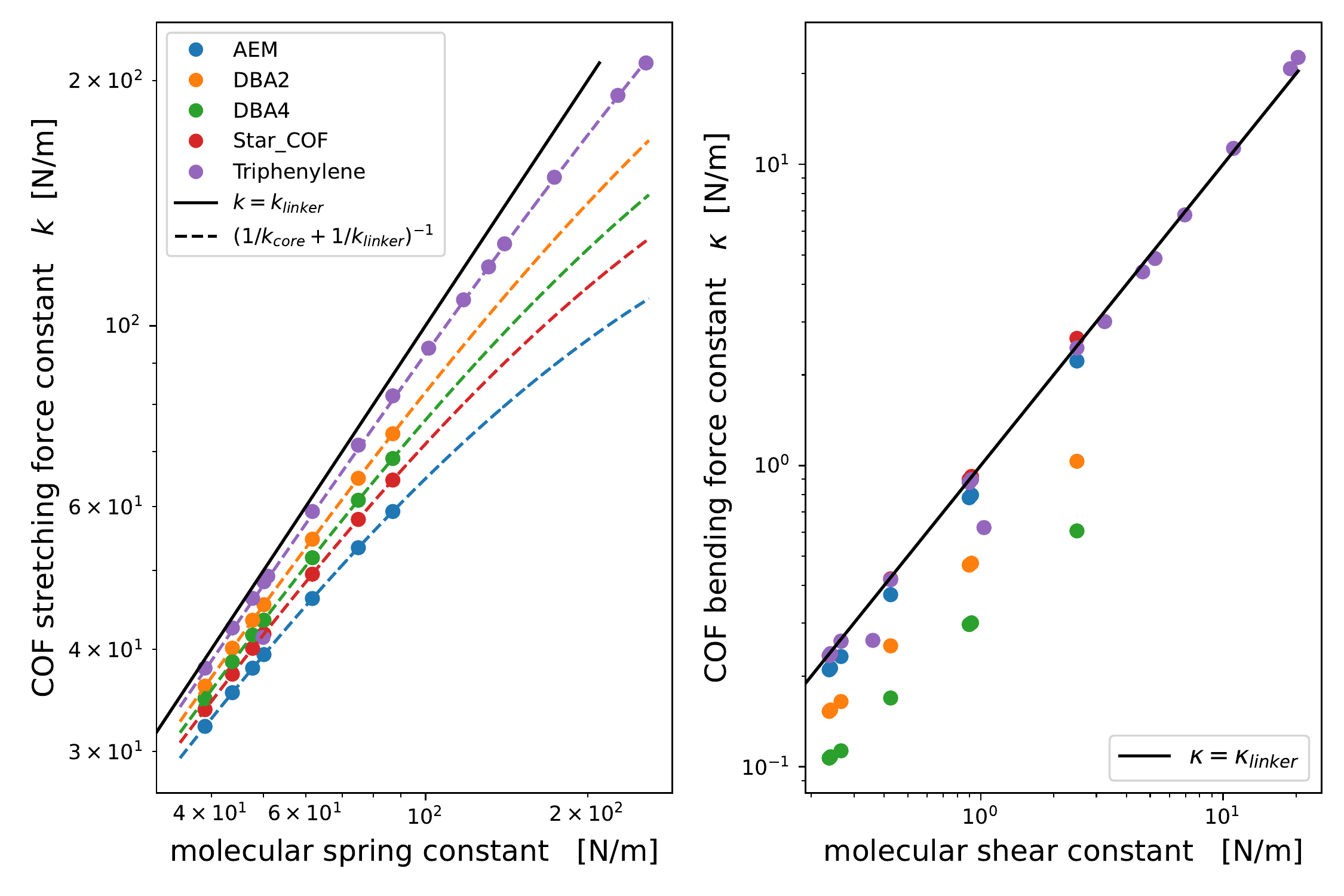}
    \caption {COF force constants vs molecular force constants for stretching (left) and bending (right). The solid lines indicates a perfect agreement. The dashed lines in the left panel are obtained by fitting Eq.\ \eref{eq:eff_spring_core} to the data for each core.}
	\label{fig:fcs_vs_fcs}
\end{figure}

The overall agreement between the molecular and the COF force-constants is very good for the COFs with Tp cores, as one can see from Fig.\ \ref{fig:fcs_vs_fcs}. Otherwise, the stretching force-constants of the COFs are overestimated by the molecular spring constant alone. The reason for this behavior can be found in neglecting the elasticity of
the cores in the model. Taking this into account, e.g.\ via a second spring representing the core, provides an improved correspondence \cite{Raptakis2021}. The effective spring constant is given by
\begin{equation}\label{eq:eff_spring_core}
    k = \left( \frac{1}{k_{\rm core}} + \frac{1}{k_{\rm linker}}\right)^{-1}\;,
\end{equation}
where $k_{\rm core}$ denotes the spring constant of the core. Using this equation we can find the respective values of $k_{\rm core}$ by fitting to the data shown in Fig.\ \ref{fig:fcs_vs_fcs}. As expected, we find the largest spring constant for the Tp cores ($k_{\rm core}\approx 1200$ N/m), while the smallest one is obtained for the AEM cores ($k_{\rm core}\approx 185$ N/m).

For the bending force-constants of the COFs we observe an almost perfect agreement for all cores except DBA2 and DBA4. Here, the non-rigidity of the core plays an important role. Nevertheless, in all cases the COF bending constant increases with increasing molecular shear constant.

\subsection{Deformation energy of defective COFs}

In order to assess the transferability of the models put forward in Secs.\ \ref{sec:spring_network} and \ref{sec:semi_flex_network} to other situations, we considered COFs with a single Stone-Wales defect. We chose this type of defect, since the number of atoms remains unchanged, which facilitates the calculation of the deformation energy. For the latter we optimized the geometry of the structures without and with defect and subtract the respective energies, i.e.\ $E_{\rm def} = E_{\rm SW} - E_{\rm clean}$. Additionally, we minimized the energies given by Eqs.\ \eref{eq:cg_energy_honeycomb} and \eref{eq:cg_energy_Mikado} using the effective stretching and bending force constants, $k$ and $\kappa$, and the lattice constant $a$ obtained from the DFTB results (table \ref{tab:moduli}).

\begin{figure*}[t!]
    \centering
    \includegraphics[width=1.\textwidth]{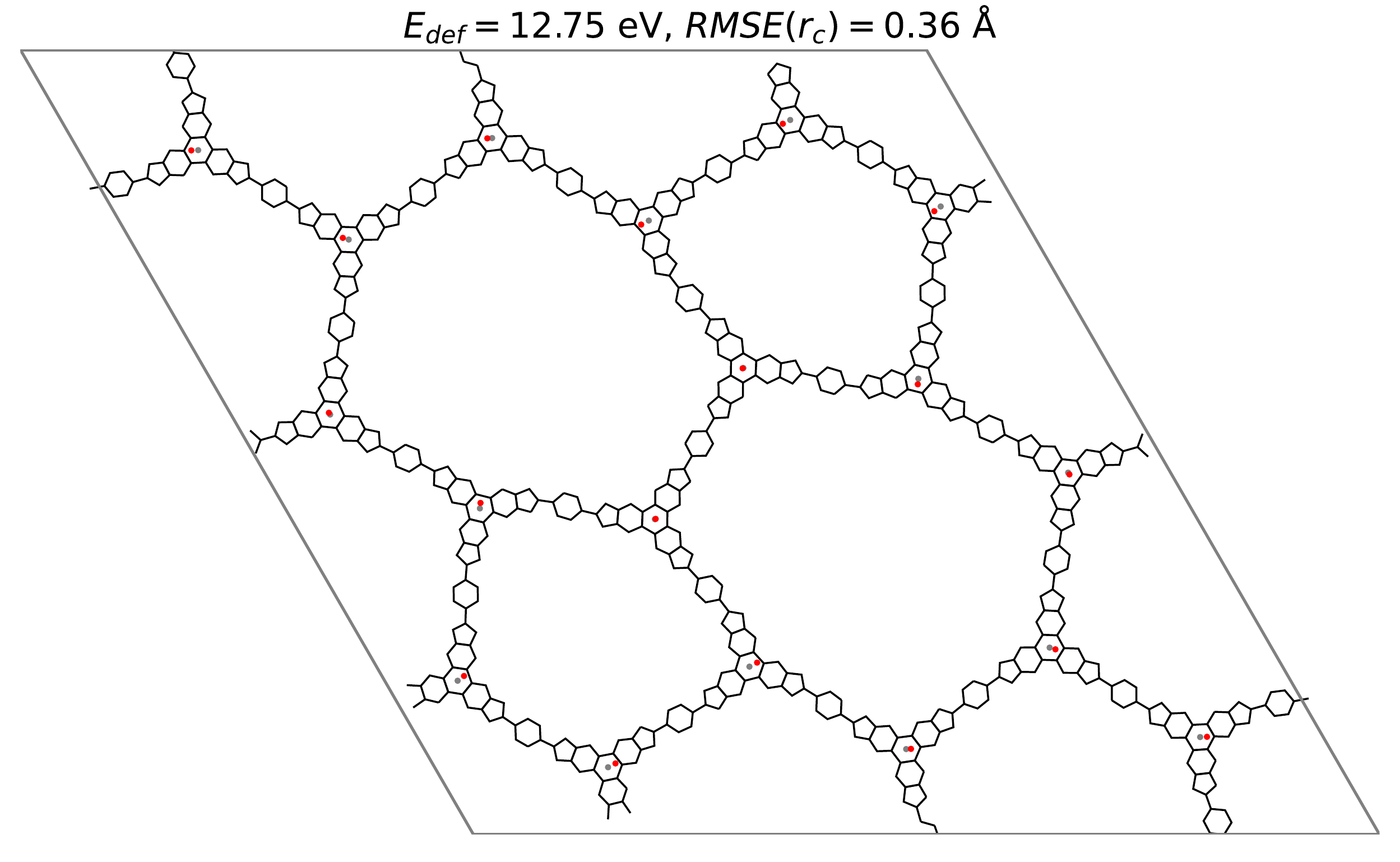}
    \includegraphics[width=1.\textwidth]{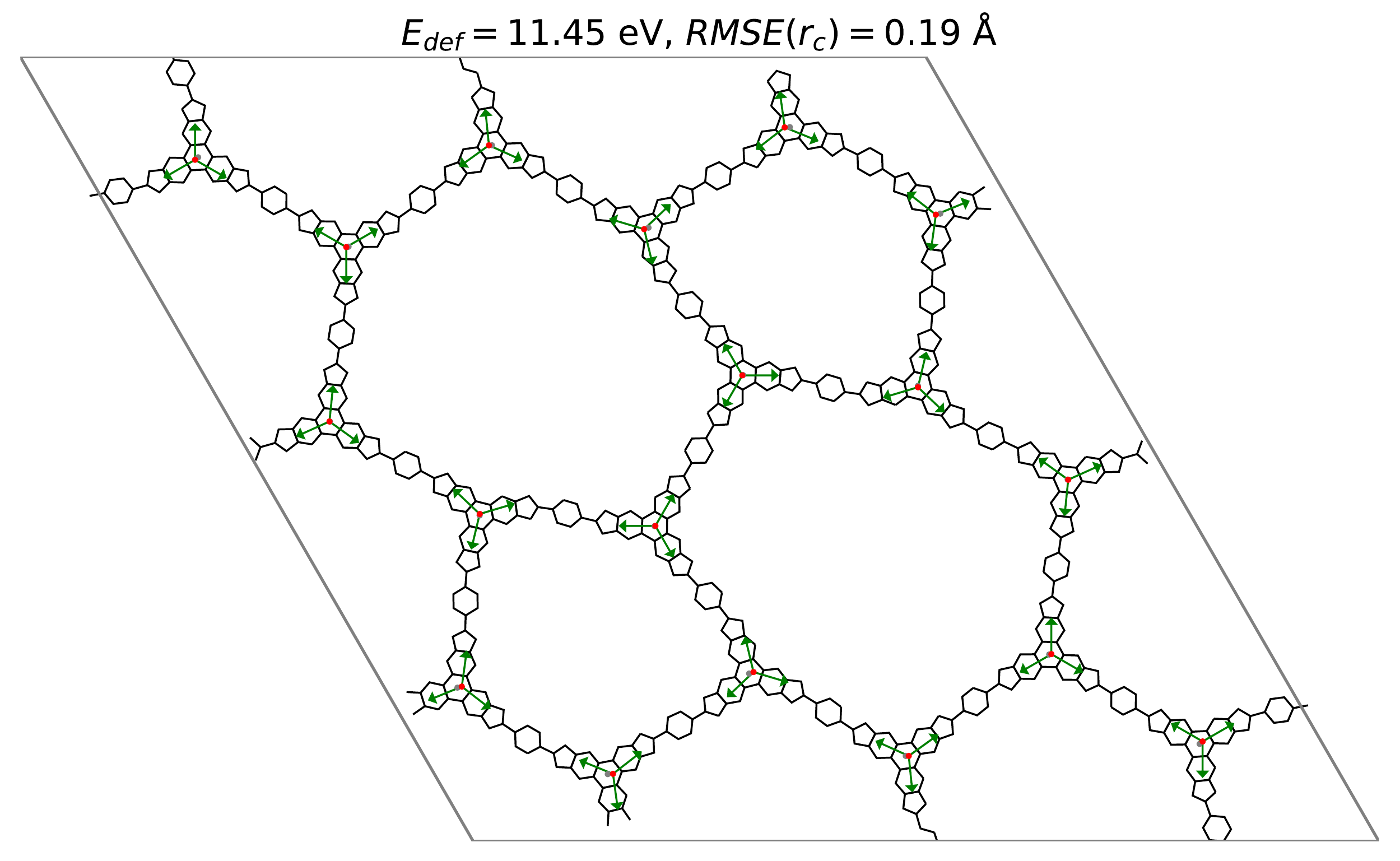}    
    \caption {Stone-Wales defect in a Tp--DB 1phenyl--COF. (top) Optimized geometry obtained from DFTB (black lines) and positions of cores from BAFF (red dots). (bottom) Optimized geometry obtained from DFTB (black lines) and positions of cores from the Mikado model (red dots). The green arrows indicate the orientation of the cores.}
	\label{fig:SW_Tp_DB}
\end{figure*}

Figure \ref{fig:SW_Tp_DB} shows a Tp--DB 1phenyl--COF with a Stone-Wales defect. The atomistic structure is schematically shown by black lines and the simulation cell is indicated by grey lines. The deformation energy in DFTB is found to be $E_{\rm def} = 12.66$ eV. This value is in close agreement with the values found by the model calculations which yield $E_{\rm def}=11.45$ eV (Mikado) and $E_{\rm def}=12.75$ eV (BAFF), respectively. The optimized positions of the core centers are indicated by red dots. The overall structure is well described by both coarse-grain models as indicated by a root-mean-square deviation of less than $0.5$ Ångstrom between the core-center coordinates of the models and the DFTB structure. The Mikado model is also able to capture the orientation of the cores and can thus provide information about the shape of the linkers.

The same procedure was repeated for the same linker but two other cores which are less rigid. Specifically, we used a Star-COF and a DBA2-COF with Stone-Wales defects (see supporting information). The deformation energies for the former COF are $E_{\rm def}^{\rm Star} = 24.06$ eV (DFTB), $E_{\rm def}^{\rm Star} = 19.8$ eV (Mikado), and $E_{\rm def}^{\rm Star} = 22.25$ eV (BAFF). For the latter COF we obtain $E_{\rm def}^{\rm DBA2} = 11.27$ eV (DFTB), $E_{\rm def}^{\rm DBA2} = 7.32$ eV (Mikado), and $E_{\rm def}^{\rm DBA2} = 8.02$ eV (BAFF). As one can see, the deformation energy is always underestimated by both models. The biggest difference to the DFTB results is found for the DBA2 COF, which has the smallest shear modulus of the three frameworks. The deformation energies in the Mikado model show the largest deviations from the DFTB results, but at the same time lead to structures with less deviation of the core positions.

\section{Conclusions}
In summary, we have calculated the in-plane elasticity of different 2D COFs with honeycomb topologies. We used DFTB to obtain the 2D bulk and shear moduli for each structure and found that both quantities decrease with increasing lattice constant. Converting the elastic constants to bond-stretching and angle-bending force-constants of a coarse-grained model, we found that the bond-stretching constants are inversely proportional to the length of the linkers, $\ell_0^{-1}$, while the angle-bending constants decrease as $\ell_0^{-3}$ or $\ell_0^{-4}$. These findings can be reproduced by modelling the individual linker molecules as 2D elastic beams like in the Mikado model. We computed the corresponding stretching and shear force-constants of the linker molecules and found a very good agreement with the COF force-constants for COFs with stiff cores. The elastic properties of the COFs can thus be predicted from the molecular building blocks. Moreover, the coarse-grained description can be used to investigate structures and phenomena on larger length-scales. As a demonstration, we considered three COFs with a Stone-Wales defect and compared the structure and deformation energy obtained within the coarse-grain models with the DFTB results. We find an overall good agreement, although the deformation energy is typically underestimated by the models.

It should be noted, that for a complete description of the elasticity of COFs additional information about the out-of-plane bending is required. For systems with many atoms in the unit-cell, the reliable calculation of the associated bending rigidities is quite challenging \cite{Kumar2020}. However, the coarse-graining approach presented here can also be applied in that case. Having the possibility to quantitatively relate the molecular properties of the building-blocks to the elastic properties of the COFs will facilitate the design and synthesis of 2D COFs with tailored properties.

\ack
A.~R.\ thanks the International Max Planck Research School "Many Particle Systems in Structured Environment" and the Chair of Materials Science and Nanotechnology in the Faculty of Mechanical Engineering of TU Dresden for financial support. D.~B.\ acknowledges financial support from the DFG project ``Chemistry of Synthetic Two-Dimensional Materials'' CRC-1415 (No.\ 417590517). We also acknowledge the Center for Information Services and High Performance Computing (ZIH) at TU Dresden for computational resources.

\bibliographystyle{iopart-num} 
\bibliography{bib_main} 

\end{document}


\title[SI for Coarse-grained elasticity of 2D COFs]{Supplementary Information for ``Towards coarse-grained elasticity of single-layer Covalent Organic Frameworks''}

\author{Alexander Croy$^{1}$, Antonios Raptakis$^{2,3}$, David Bodesheim$^{2}$, Arezoo Dianat$^{2}$, and Gianaurelio Cuniberti$^{2,4}$}
\address{$^{1}\space $ Institute of Physical Chemistry, Friedrich Schiller University Jena, 07737 Jena, Germany}
\address{$^{2}\space $ Institute for Materials Science and Max Bergmann Center of Biomaterials, TU Dresden, 01062 Dresden, Germany}
\address{$^{3}\space $ Max Planck Institute for the Physics of Complex Systems, 01187 Dresden, Germany}
\address{$^{4}\space $ Dresden Center for Computational Materials Science (DCMS), TU Dresden, 01062 Dresden, Germany}
\ead{alexander.croy@uni-jena.de}

\section{In-plane elasticity of isotropic 2D materials}
The energy density (energy $E$ per area $A_0$) of an orthotropic 2D material can be written as \cite{ma19}:
\begin{eqnarray}\label{eq:ortho_dens}
    \mathcal{E} &={}& \frac{E}{A_0} = \frac{1}{2}
    \left(\begin{array}{c}
        \varepsilon_{xx} \\
        \varepsilon_{yy} \\
        2\varepsilon_{xy}
    \end{array}\right)^T
    \left(\begin{array}{ccc}
        C_{11} & C_{12} & 0 \\
        C_{12} & C_{22} & 0 \\
        0 & 0 & C_{66}
    \end{array}\right)
    \left(\begin{array}{c}
        \varepsilon_{xx} \\
        \varepsilon_{yy} \\
        2\varepsilon_{xy}
    \end{array}\right) \\
    &={}& \frac{1}{2}\left(
     C_{11} \varepsilon_{xx}^2 + C_{22} \varepsilon_{yy}^2
     + 2 C_{12} \varepsilon_{xx} \varepsilon_{yy}
     + 4 C_{66}\varepsilon_{xy}^2
    \right)\;, \nonumber
\end{eqnarray}
where $\varepsilon_{ij}$ denote the elements of the symmetric strain tensor and $C_{ij}$ are the elastic constants in Voigt notation.

For an isotropic material the energy density depends only on \textit{invariants} of the strain tensor, i.e.\ its eigenvalues. Within linear elasticity, only quadratic terms of the eigenvalues can appear and typically the energy density is then written in terms of the square of the trace and the trace of the square of $\bm{\varepsilon}$ \cite{lali86}:
\begin{eqnarray}\label{eq:iso_dens}
    \mathcal{E} &={}& \frac{1}{2} \lambda_{2D} \left(\Tr \bm{\varepsilon}\right)^2
     + \mu_{2D} \Tr \bm{\varepsilon}^2 \\
     &={}&\frac{1}{2}\left[
        \lambda_{2D} (\varepsilon_{xx} + \varepsilon_{yy})^2
         + 2\mu_{2D} (\varepsilon_{xx}^2  + \varepsilon_{yy}^2 + 2 \varepsilon_{xy}^2)\right]\;. \nonumber
\end{eqnarray}
Here, $\lambda_{2D}$ and $\mu_{2D}$ are the material-specific 2D Lam\'e coefficients. Setting Eqs.\ \eref{eq:ortho_dens} and \eref{eq:iso_dens} equal, one sees that
\begin{eqnarray*}
    \mathcal{E} &={}& \frac{1}{2}\left[
        (\lambda_{2D} + 2\mu_{2D}) (\varepsilon_{xx}^2 + \varepsilon_{yy}^2)
         + 2\lambda_{2D} \varepsilon_{xx} \varepsilon_{yy} + 4\mu_{2D} \varepsilon_{xy}^2\right] \\
    &={}&
         \frac{1}{2}\left[
     (C_{11}\varepsilon_{xx}^2 + C_{22}\varepsilon_{yy}^2)
     + 2 C_{12} \varepsilon_{xx} \varepsilon_{yy}
     + 4 C_{66}\varepsilon_{xy}^2
    \right]\;,
\end{eqnarray*}
which implies
\begin{equation}\label{eq:iso_constants}
    C_{11} = C_{22} =\lambda_{2D} + 2\mu_{2D},\quad
    C_{12} = \lambda_{2D},\quad
    C_{66} = \mu_{2D} = \frac{1}{2} \left(  C_{11} - C_{12} \right)\;.
\end{equation}

Focusing on the isotropic case we will consider two types of biaxial deformations: a) uniform stretching/compression ($\varepsilon_{xx}=\varepsilon_{yy}=\varepsilon/2$, $\varepsilon_{xy}=0$) and b) shearing ($\varepsilon_{xx}=-\varepsilon_{yy}=\varepsilon/2$, $\varepsilon_{xy}=0$). Inserting the strain tensor components into the energy density yields
\begin{eqnarray}
    \mathcal{E}&={}&
         \frac{1}{2}\left[
    C_{11} (\varepsilon^2 + \varepsilon^2)/4
     \pm 2 C_{12} \varepsilon^2/4
    \right]
    = \frac{1}{2} \left(\frac{
    C_{11} \pm C_{12} }{2} \right)
     \varepsilon^2\;.
\end{eqnarray}
The expression in brackets corresponds to the bulk modulus (case a) and the shear modulus (case b). The latter was already found in Eq.\ \eref{eq:iso_constants} and the former can be verified by considering the definition of the bulk modulus,
\begin{equation}
    B_{2D} =  \frac{C_{11} + C_{12}}{2}, \quad
    \mu_{2D}= \frac{C_{11} - C_{12}}{2}\;.
\end{equation}

Lastly, we consider uniaxial tension/compression along the $x$-direction. The stress tensor components $\sigma_{ij}$ are given by Hooke's law,
\begin{eqnarray}\label{eq:Hooke}
    \left(\begin{array}{c}
        \sigma_{xx} \\
        \sigma_{yy} \\
        \sigma_{xy}
    \end{array}\right)
    &={}&
    \left(\begin{array}{ccc}
        C_{11} & C_{12} & 0 \\
        C_{12} & C_{11} & 0 \\
        0 & 0 & \left(  C_{11} - C_{12} \right)/2
    \end{array}\right)
    \left(\begin{array}{c}
        \varepsilon_{xx} \\
        \varepsilon_{yy} \\
        2\varepsilon_{xy}
    \end{array}\right) \;.
\end{eqnarray}
For the uniaxial deformation $\sigma_{yy}=\sigma_{xy}=0$ and solving the resulting system of equations for $\sigma_{xx}$ and $\epsilon_{yy}$ yields
\begin{equation}
    \sigma_{xx} =  \frac{C_{11}^2 -C_{12}^2}{C_{11}} \epsilon_{xx} = Y_{2D} \epsilon_{xx}, \quad
    \epsilon_{yy}= - \frac{C_{12}}{C_{11}} \epsilon_{xx}
    = - \nu_{2D} \epsilon_{xx}\;.
\end{equation}
The first expression defines the Young modulus and the second the Poisson ratio,
\begin{equation}
    Y_{2D} =  \frac{C_{11}^2 -C_{12}^2}{C_{11}} = \frac{4 B_{2D} \mu_{2D}}{B_{2D} + \mu_{2D}}, \quad
    \nu_{2D}= \frac{C_{12}}{C_{11}} = \frac{B_{2D} - \mu_{2D}}{B_{2D} + \mu_{2D}}\;.
\end{equation}
Note that the 2D Poisson ratio can take values between $-1$ and $1$, while the 3D Poisson ratio ranges between $-1$ and $1/2$.

\section{Details of the Mikado model}
As stated in the main text, the individual fibers in the Mikado model are described as elastic beams and their energy under deformation is given by
\begin{equation}\label{eq:E_fiber}
    \mathcal{H}_{\rm fiber} = \int_0^{\ell_0} ds \left\{ \frac{c_{\rm stretch}}{2} \left(\frac{\partial u_{\|}(s)}{\partial s}\right)^2 + 
    \frac{c_{\rm bend}}{2}\left(\frac{\partial^2 u_{\perp}(s)}{\partial s^2}\right)^2\right\}\;.
\end{equation}
To determine the equilibrium tangential and transverse displacements, $u_{\|}(s)$ and $u_{\perp}(s)$, one considers the variation of the energy. This yields the following differential equations
\begin{eqnarray}
    \frac{d^2}{ds^2}u_{\|}(s) &=& 0 \;,\\
    \frac{d^4}{ds^4}u_{\perp}(s) &=& 0 \;,
\end{eqnarray}
which have to be supplemented with the boundary conditions $u_{\|}(0), u_{\|}(\ell_0)$ and $u_{\perp}(0), u_{\perp}(\ell_0), d u_{\perp}/ds(0), d u_{\perp}/ds(\ell_0)$. The solutions of the differential equations are polynomials of degree one and three, respectively, and the coefficients are given in terms of the boundary conditions:
\begin{eqnarray}
    u_{\|}(s) ={}& \left(\frac{u_{\|}(\ell_0)-u_{\|}(0)}{\ell_0}\right) s + u_{\|}(0) \label{eq:u_par}\\
    u_{\perp}(s) ={}& \left( \frac{\varphi(0) + \varphi(\ell_0)}{\ell_0^2} - 2\frac{ u_\perp(\ell_0)-u_\perp(0)}{\ell_0^3}\right) s^3 \nonumber\\ 
    {}&+ \left( -\frac{2\varphi(0) + \varphi(\ell_0)}{\ell_0} + 3\frac{ u_\perp(\ell_0)-u_\perp(0)}{\ell_0^2}\right) s^2 \nonumber\\
    {}& + \varphi(0) s + u_\perp(0)\;, \label{eq:u_perp}
\end{eqnarray}
where $\varphi(s) = d u_{\perp}/ds(s)$. By inserting the expressions \eref{eq:u_par} and \eref{eq:u_perp} into the deformation energy \eref{eq:E_fiber} one readily finds the stretching and bending contributions to $\mathcal{H}_{\rm fiber}$ as stated in the main text.

\begin{figure*}[b!]
    \centering
    \includegraphics[width=0.8\textwidth]{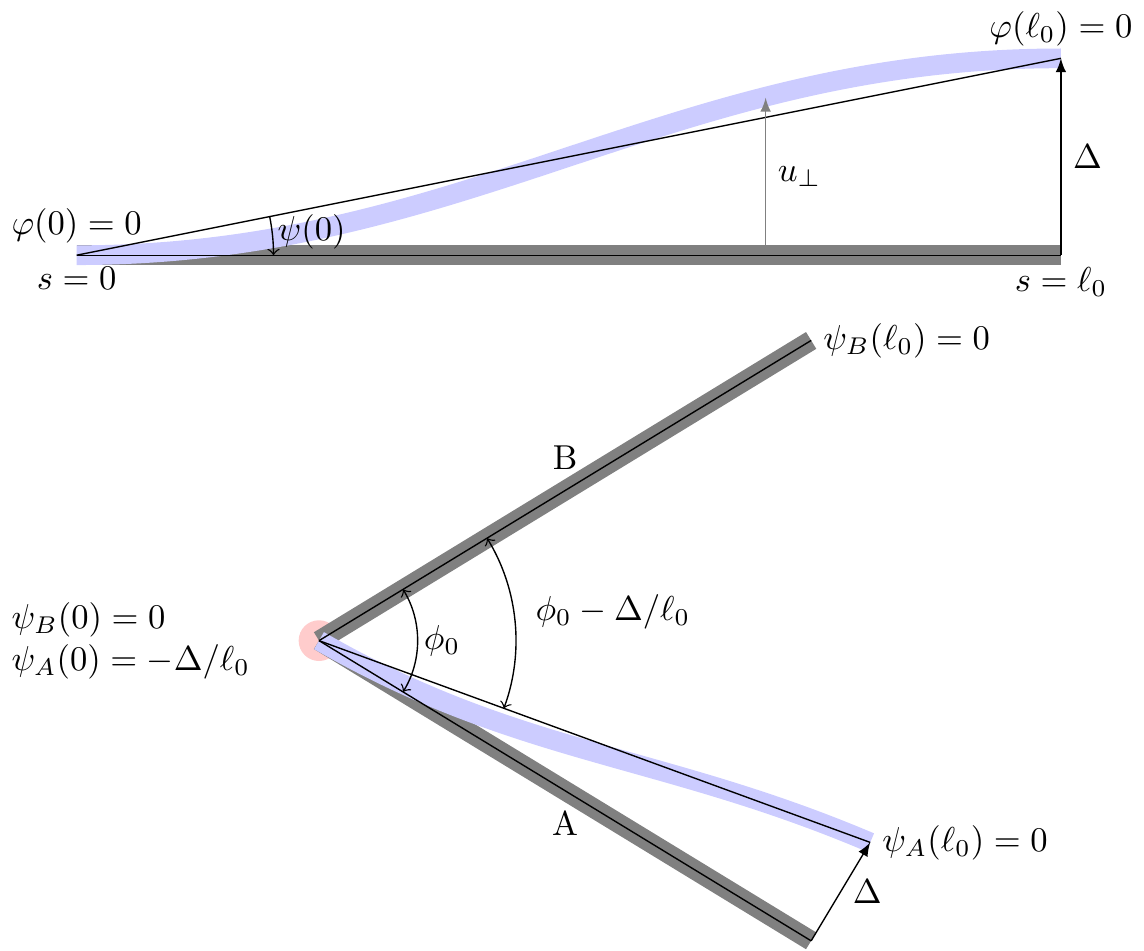}
    \caption {Bending of fibers in the Mikado model. (top) Displacing one end of a fiber perpendicular to the fiber axis. (bottom) Two fibers linked at a common joint while one of them is displaced at the end. }
	\label{fig:Mikado_bending}
\end{figure*}

According to Eq.\ \eref{eq:u_par}, stretching an individual fiber leads to a homogeneous deformation with constant strain $d u_{\|}/ds$ along the fiber. For bending, the situation is more complicated due to the interplay of the boundary conditions. In the upper part of Fig.\ \ref{fig:Mikado_bending}, the situation is sketched where one end of a fiber is displaced perpendicular to the fiber axis while keeping the slopes $\varphi(s)$ at the ends fixed. If the latter are also allowed to change, the fiber returns to a straight configuration. If two fibers are linked at a rigid joint, the slopes are constrained and the fibers are necessarily bent when one end is displaced. This is illustrated in the lower part of Fig.\ \ref{fig:Mikado_bending}.

\begin{figure*}[b!]
    \centering
    \includegraphics[width=\textwidth]{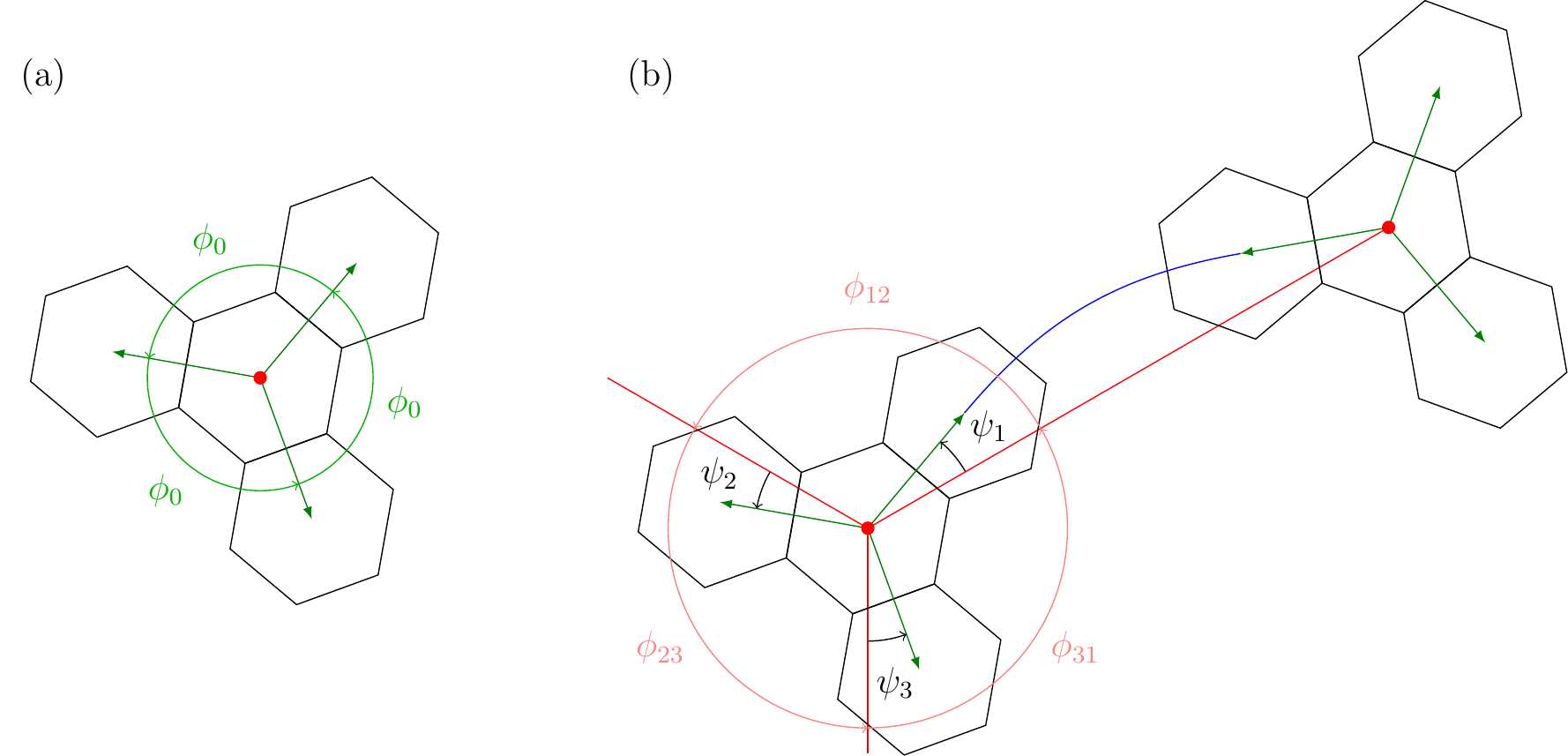}
    \caption {(a) In the Mikado model each core is assumed to be a rigid molecule specified by its position (red dot) and an orientation (one of the green arrows). The angle between the three linking sites is constrained to $\phi_0=2\pi/3$. (b) The deformation of the linkers (indicated as blue line) is determined by the distance of the cores (red lines) and the orientation of the cores (angles $\psi_i$) relative to the distance vectors. }
	\label{fig:Mikado_joints}
\end{figure*}

To obtain the position and the orientation of a core molecule, we use the center of the coordinates of the molecule and of one of the ``linking phenyls'', respectively. In Fig.\ \ref{fig:Mikado_joints}(a) this is illustrated for the Tp core. Since the angle between the resulting three orientation vectors is fixed, we can choose any one of them as a reference and label them clockwise as $i=1,2,3$. Within a cross-linked network, we find the vectors connecting the core to its neighbors and label them clockwise as well. Then we compute the angles $\psi_i$ between the connection vectors and the associated orientation vector as shown in Fig.\ \ref{fig:Mikado_joints}(b). For a flat structure, we have $\phi_{12}+\phi_{23}+\phi_{31}=2\pi$ and we can express $\psi_2$ and $\psi_3$ in terms of $\psi_1$:
\begin{eqnarray*}
    \psi_2 =& \psi_1 + \phi_0 - \phi_{12}\;,\\
    \psi_3 =& \psi_1 + 2\phi_0 - (\phi_{12} + \phi_{23})\;, \\
    \psi_1 =& \psi_1 + 3\phi_0 - (\phi_{12} + \phi_{23}+\phi_{31})
    = \psi_1\;.
\end{eqnarray*}

\section{Linker-deformation in COFs}
\begin{figure*}[t!]
    \centering
    \includegraphics[width=7cm]{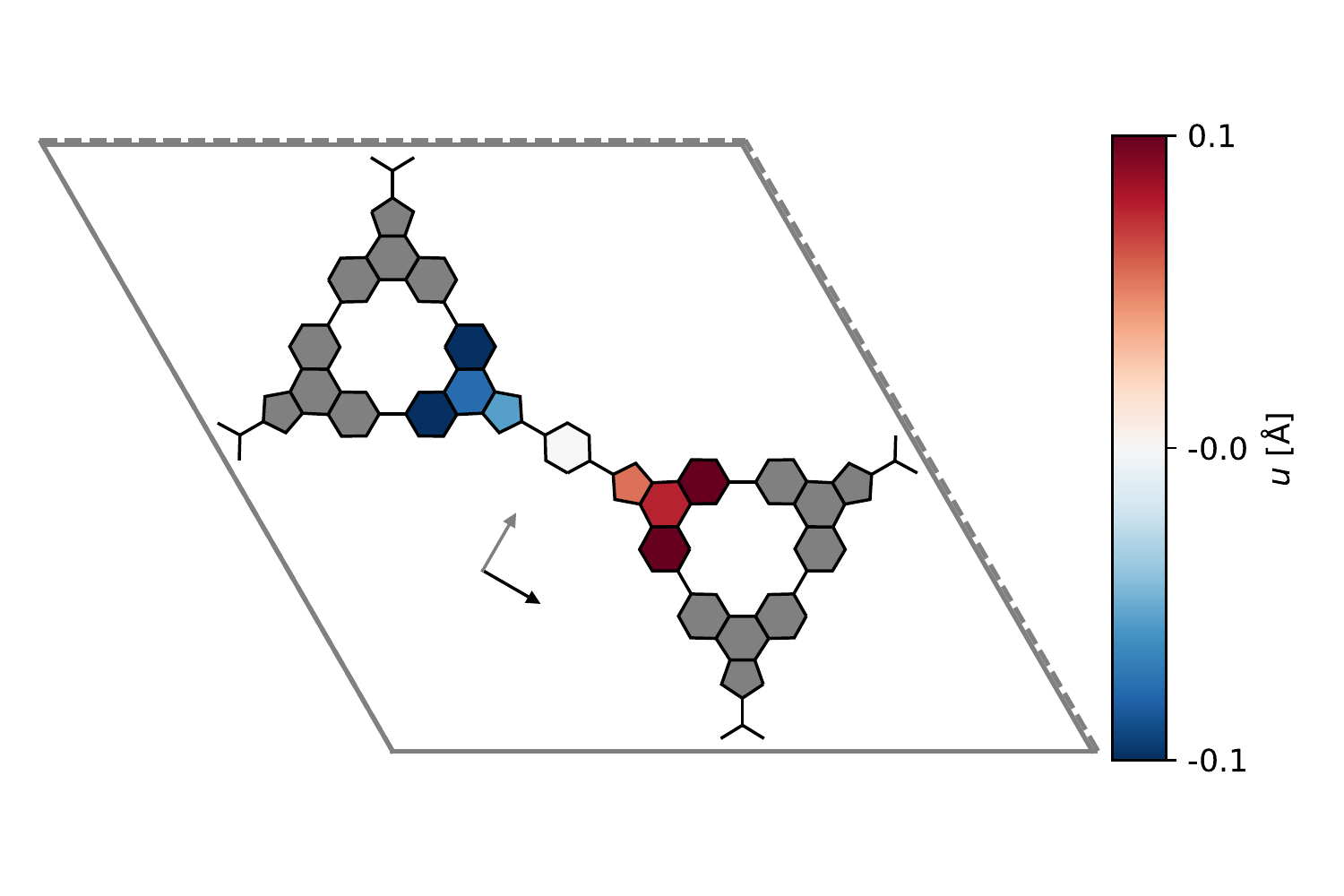}
    \includegraphics[width=7cm]{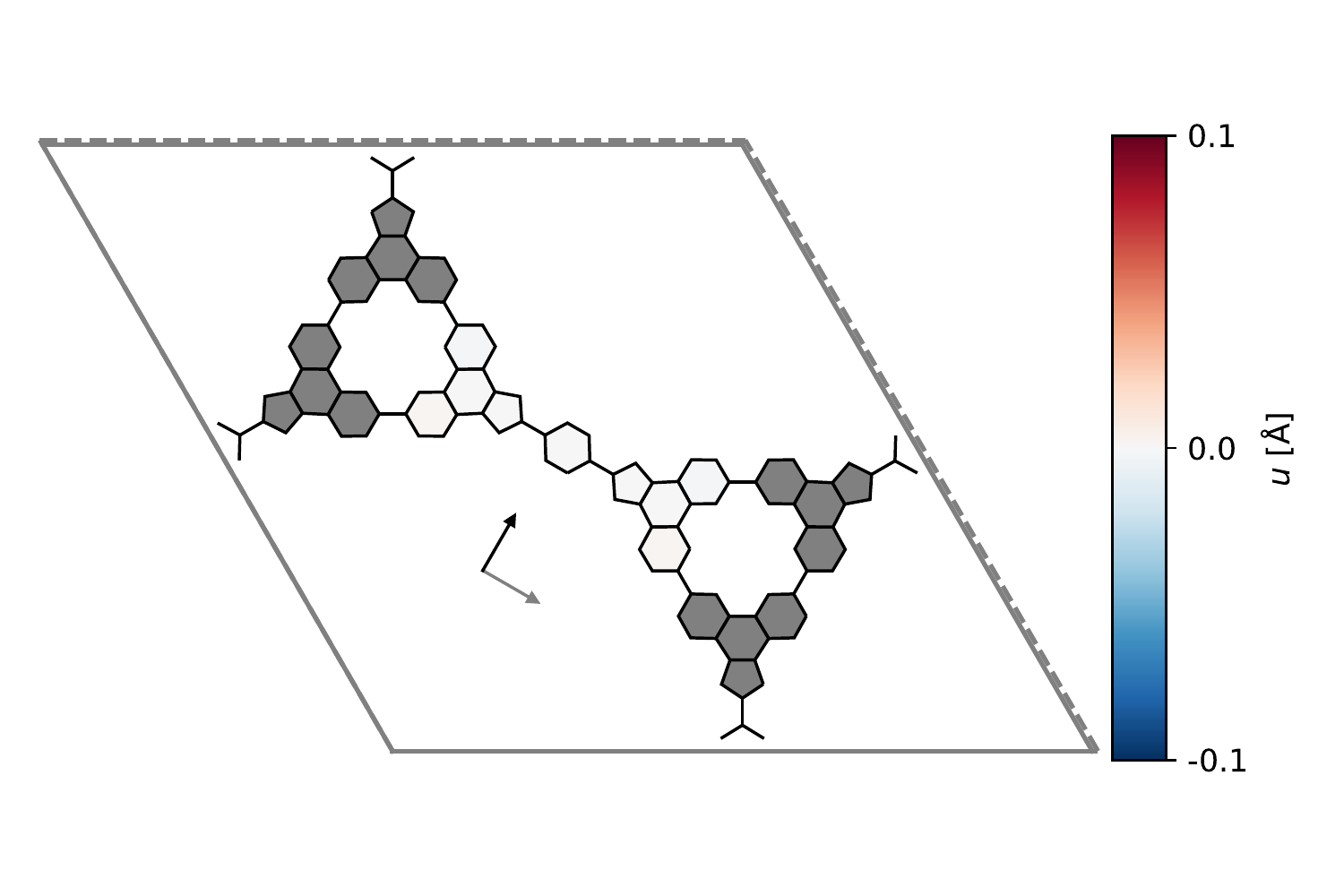}\\
    \includegraphics[width=7cm]{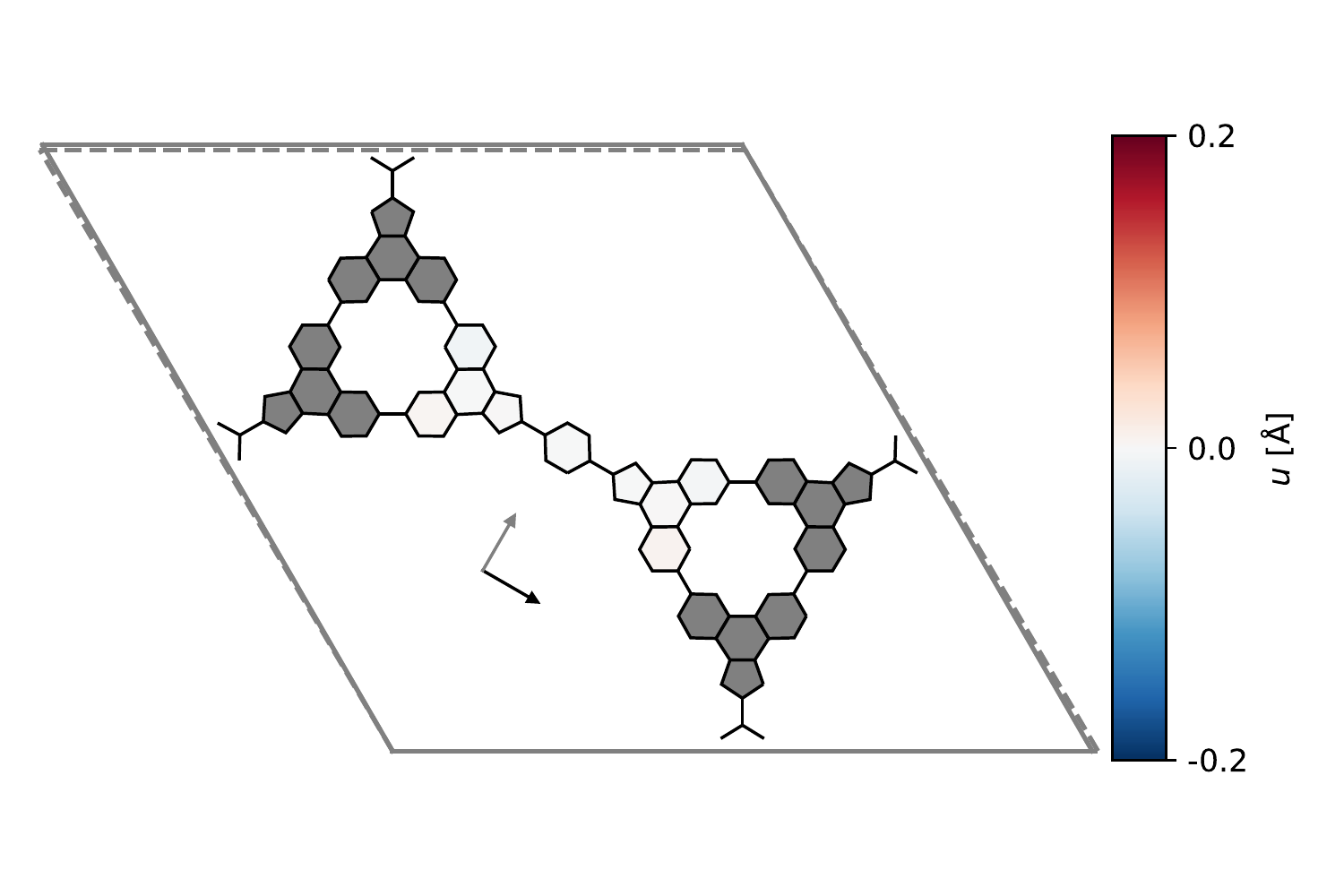}
    \includegraphics[width=7cm]{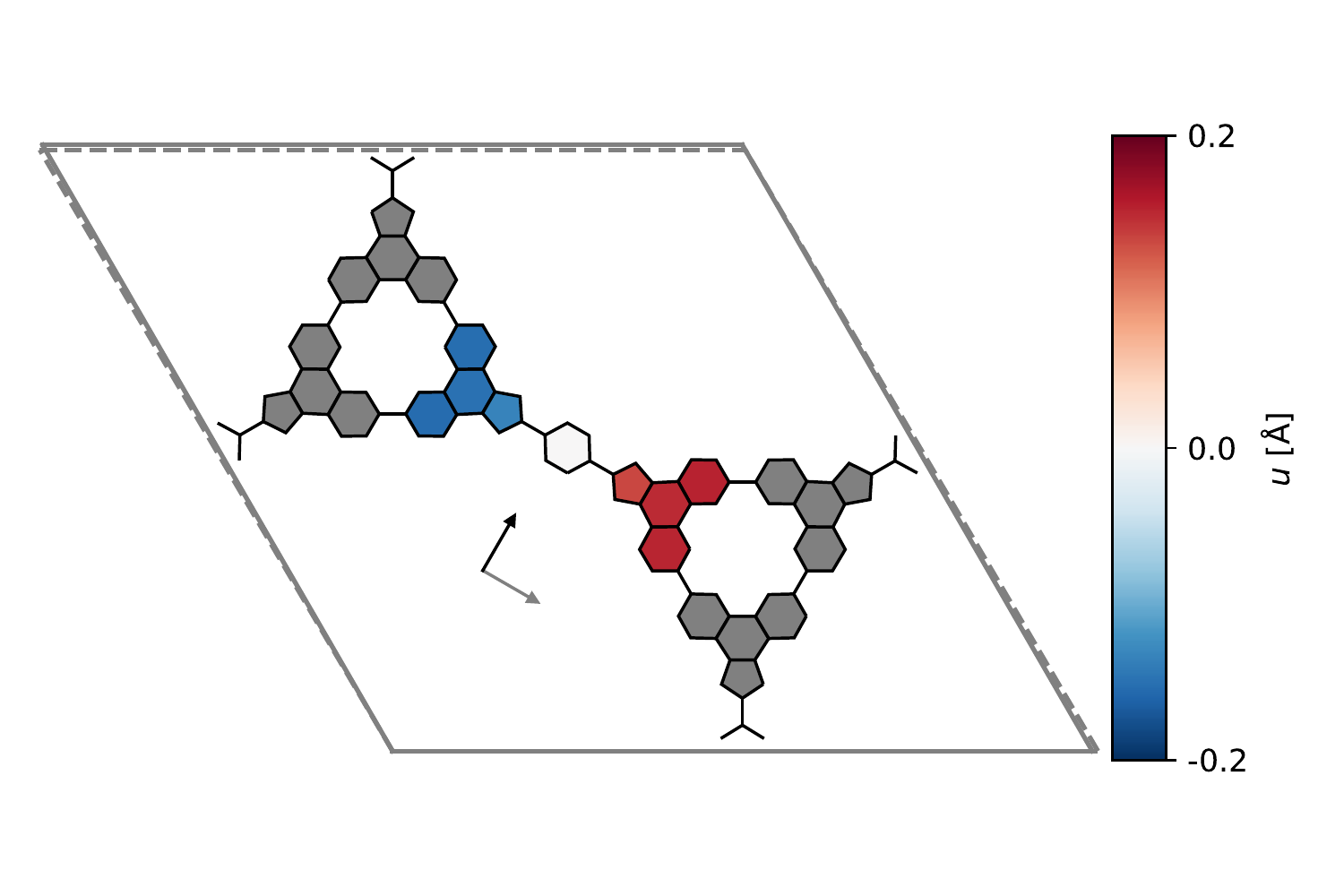}
    \caption {Displacement for one selected linker in a deformed Star-COF. The upper and lower row shows uniaxial stretching (by $0.8\%$) and shearing (by $0.8\%$), respectively. The unit cell is indicated by solid (dashed) lines for the undeformed (deformed) structure. For the visualization we average the displacement of atoms within each ring (hexagon or pentagon) and subtract the average displacement of the linker. The displacements are projected on the directions parallel (left) and perpendicular (right) to the linker axis in the undeformed state as indicated by the arrows. For convenience the displacement of only one linker is shown, the remaining COF is shown in grey.}
	\label{fig:COF_deformation}
\end{figure*}

As an example for the linker-deformation, in Fig.\ \ref{fig:COF_deformation}
we show the displacement within one linker of a Star-COF projected on the directions parallel and perpendicular to its axis in the undeformed state, which is indicated by the arrows. For the visualization we average the displacement of atoms within each ring (hexagon or pentagon) and subtract the average displacement of the linker. As one can clearly see in case of the uniaxial deformation (upper row), the displacement is mostly in the direction of the molecular axis (left) and there is only negligible displacement in the perpendicular direction (right). For the shearing (lower row), one observes the opposite behavior.

\section{Stone-Wales defect in DBA2 and Star COFs}
\begin{figure*}[t!]
    \centering
    \includegraphics[width=1.\textwidth]{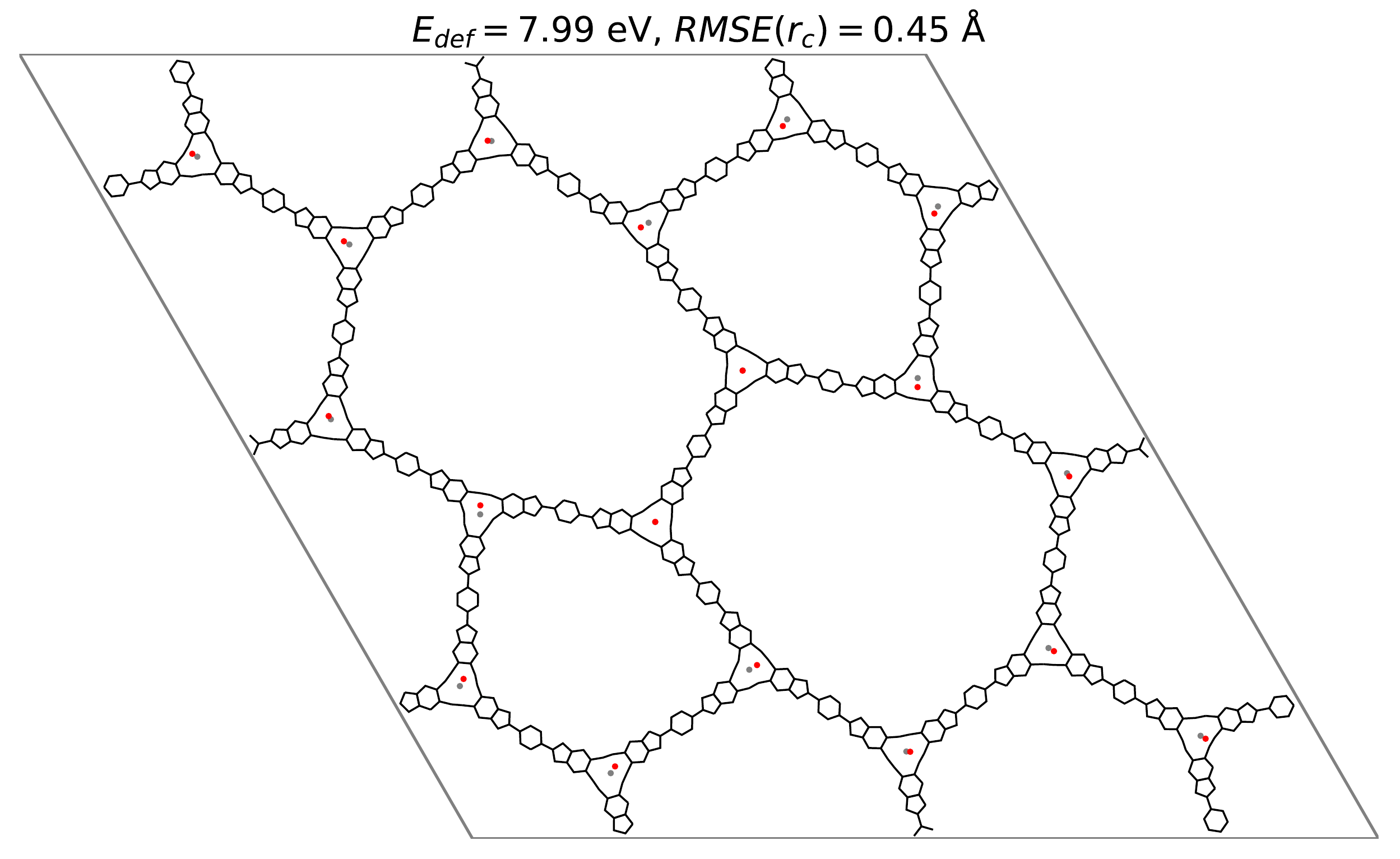}
    \includegraphics[width=1.\textwidth]{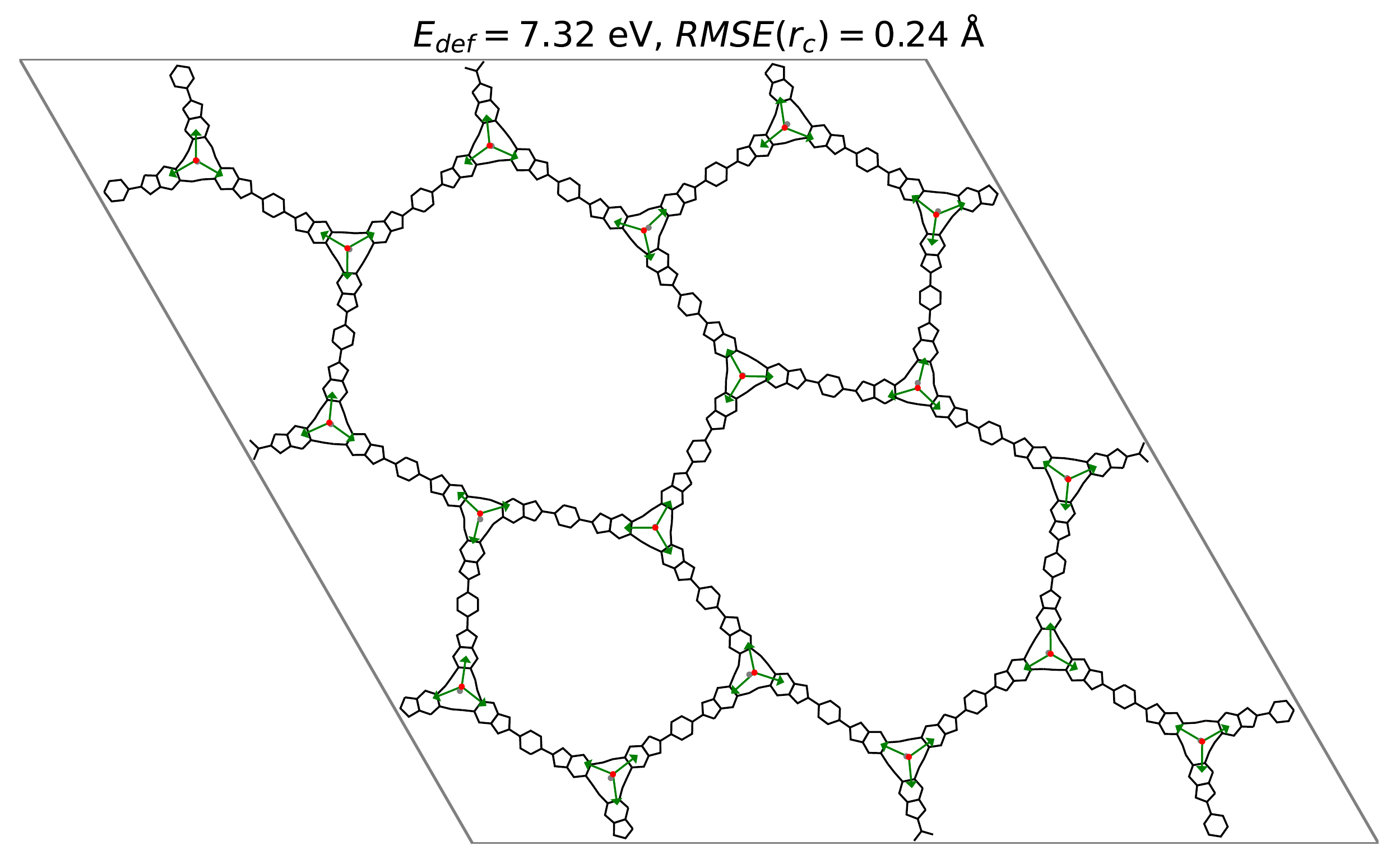}    
    \caption {Stone-Wales defect in a DBA2--1phenyl--COF. (top) Optimized geometry (black lines, grey dots) and positions of cores (red dots) obtained from DFTB and the BAFF model, respectively. (bottom) Optimized geometry (black lines, grey dots) and positions of cores (red dots) obtained from DFTB and the Mikado model, respectively. The green arrows indicate the orientation of the cores.}
	\label{fig:SW_DBA2}
\end{figure*}

\begin{figure*}[t!]
    \centering
    \includegraphics[width=1.\textwidth]{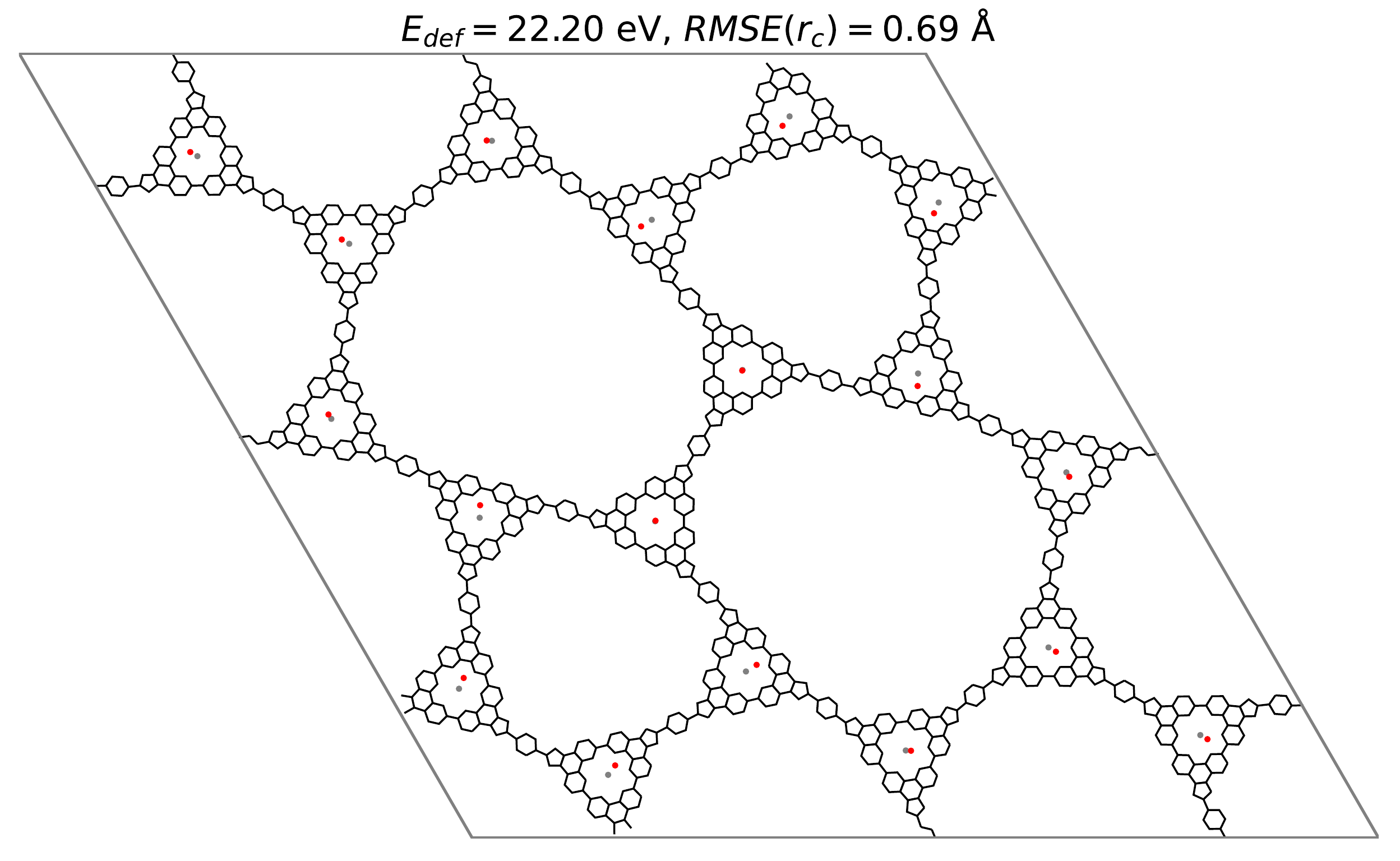}
    \includegraphics[width=1.\textwidth]{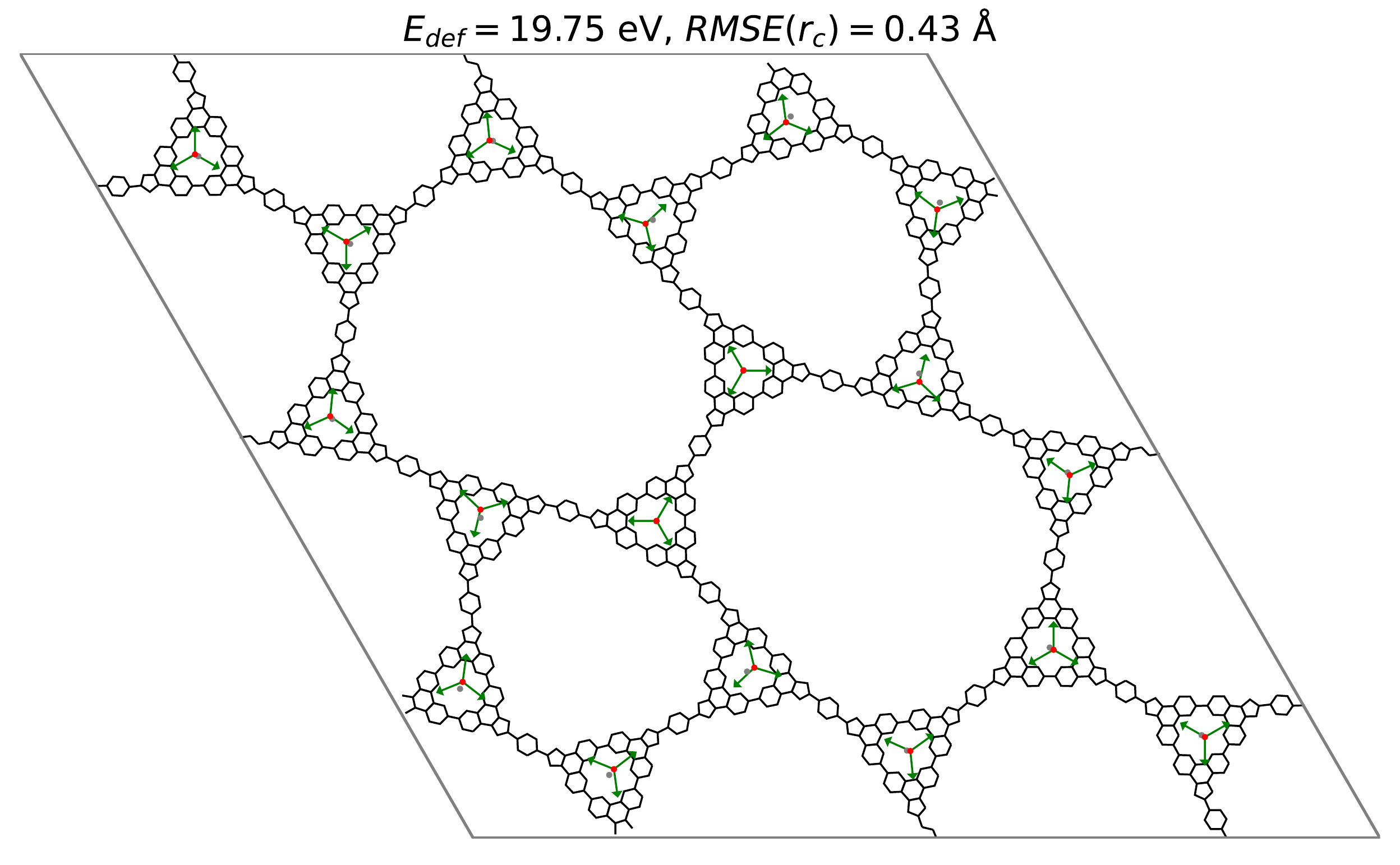}    
    \caption {Stone-Wales defect in a Star--1phenyl--COF. (top) Optimized geometry (black lines, grey dots) and positions of cores (red dots) obtained from DFTB and the BAFF model, respectively. (bottom) Optimized geometry (black lines, grey dots) and positions of cores (red dots) obtained from DFTB and the Mikado model, respectively. The green arrows indicate the orientation of the cores.}
	\label{fig:SW_Star}
\end{figure*}

Figures \ref{fig:SW_DBA2} and \ref{fig:SW_Star} show two further examples of COFs with a Stone-Wales defect. 

\bibliographystyle{iopart-num} 
\bibliography{bib_main} 